\documentclass{aa}

\usepackage{graphicx}
\usepackage[varg]{txfonts}

\newcommand{\kms}{$\mathrm{km}\,\mathrm{s}^{-1}$}
\newcommand{\aforty}{$\alpha$.40}
\newcommand{\mhi}{$M_{HI}$}
\newcommand{\Vrot}{$V_{rot,HI}$}
\newcommand{\Vhalo}{$V_{h,max}$}
\newcommand{\vrot}{$V_{rot}$}
\newcommand{\vhalo}{$V_h$}
\newcommand{\vrotvhalo}{$V_{rot}$-$V_h$}
\newcommand{\fhi}{$f_{HI}$}
\newcommand{\LCDM}{$\Lambda$CDM}

\begin{document}

\title{Is there a ``too big to fail'' problem in the field?}

\author{E. Papastergis\inst{\ref{kapteyn}} \and R. Giovanelli\inst{\ref{cornell}} \and M.P. Haynes\inst{\ref{cornell}} \and F. Shankar\inst{\ref{southampton}}}

\institute{
Kapteyn Astronomical Institute, University of Groningen, Landleven 12, Groningen NL-9747AD, Netherlands \email{papastergis@astro.rug.nl}\label{kapteyn}
\and
Astronomy Department, Space Sciences Building, Cornell University, Ithaca, NY 14853, USA \email{riccardo@astro.cornell.edu, haynes@astro.cornell.edu}\label{cornell}
\and
School of Physics and Astronomy, University of Southampton, Southampton SO17 1BJ, UK \email{f.shankar@soton.ac.uk}\label{southampton}
}

\abstract{We use the Arecibo legacy fast ALFA (ALFALFA) 21cm survey to measure the number density of galaxies as a function of their rotational velocity, \Vrot \ (as inferred from the width of their 21cm emission line). Based on the measured velocity function we statistically connect  galaxies with their host halo, via abundance matching. In a lambda cold dark matter (\LCDM) cosmology, dwarf galaxies are expected to be hosted by halos that are significantly more massive than indicated by the measured galactic velocity; if smaller halos were allowed to host galaxies, then ALFALFA would measure a much higher galactic number density. We then seek observational verification of this predicted trend by analyzing the kinematics of a literature sample of gas-rich dwarf galaxies. We find that galaxies with \Vrot$\lesssim 25$ \kms \ are kinematically incompatible with their predicted \LCDM \ host halos, in the sense that hosts are too massive to be accommodated within the measured galactic rotation curves. This issue is analogous to the ``too big to fail'' problem faced by the bright satellites of the Milky Way, but here it concerns extreme dwarf galaxies in the \textit{field}. Consequently, solutions based on satellite-specific processes are not applicable in this context. Our result confirms the findings of previous studies based on optical survey data and addresses a number of observational systematics present in these works. Furthermore, we point out the assumptions and uncertainties that could strongly affect our conclusions. We show that the two most important among them -namely baryonic effects on the abundances of halos and on the rotation curves of halos- do not seem capable of resolving the reported discrepancy.}


\maketitle

\section{Introduction}
\label{sec:intro}

The lambda cold dark matter (\LCDM) cosmological paradigm has been extremely successful in reproducing the observed expansion history and large-scale structure of the universe. Remarkably, measurements of sub-percent accuracy of the cosmic microwave background and similarly accurate measurements of the large scale distribution of galaxies have yielded no evidence of any deviations from the ``standard'' cosmological model \citep{Planck2013, Samushia2013}. However, testing \LCDM \ on galactic and subgalactic scales is considerably more difficult (both from an observational and a theoretical point of view), and several potential discrepancies between theoretical predictions of \LCDM \ and observations have been pointed out in the literature.

The ``missing satellites problem'' is perhaps the most widely known and most investigated issue. It refers to the large discrepancy between the number of low-mass subhalos expected within a Milky Way (MW)-sized halo and the number of actual MW satellites observed \citep{Klypin1999, Moore1999}. Similar discrepancies have been discovered in the field in various contexts, such as those related to the paucity of low-mass galaxies in voids \citep[``void phenomenon'';][]{Peebles2001}, the sizes of mini-voids in the Local Volume \citep{TikhonovKlypin2009}, or the slowly rising galaxy velocity function \citep{Zwaan2010, Papastergis2011,Klypin2014}. All these concerns are different aspects of the same ``overabundance problem'', whereby \LCDM \ predicts a quickly rising number of halos with decreasing halo mass, while the number of low-mass galaxies increases much more slowly.

Unfortunately, the cosmological interpretation of overabundance issues is ambiguous. For example, it is not clear whether the missing satellites problem signals a failure of \LCDM \ on small scales or if it implies that most MW subhalos remain dark and therefore undetectable. In fact, there are a number of internal baryonic processes (such as supernova and radiation pressure feedback; \citealp[e.g.,][]{Governato2010,Trujillo2013}) and environmental mechanisms (photoionization feedback, tidal and ram-pressure stripping, etc.; \citealp[e.g.,][]{Bullock2000,Somerville2002,Okamoto2008,Arraki2014,Zolotov2012,Collins2013,Kravtsov2004}) that are expected to inhibit the formation of galaxies in the smallest MW subhalos. Low-mass halos in the field would also be affected by internal baryonic processes, resulting in galaxies that are faint and therefore hard to detect in surveys (e.g., proposed solution to the void phenomenon by \citealp{TinkerConroy2009}). Even the potential challenges based on galactic rotational velocities \citep{TikhonovKlypin2009,Zwaan2010,Papastergis2011,Klypin2014} are subject to baryonic complications, which are related to the shape of the rotation curve of dwarf galaxies.

Even though number counts of low-mass galaxies do not provide a stringent test of \LCDM \ by themselves, they provide the basis for a different challenge that is much more difficult to resolve. \citet{Boylan2011} first identified the potential issue in the context of the MW satellites, and dubbed it the ``too big to fail'' (TBTF) problem. Given the low number of observed MW satellites, galaxy formation should be restricted to the few most massive subhalos of the MW. However, according to dark-matter-only (DM-only) simulations (Aquarius simulation; \citealp{Springel2008}), the likely hosts are too dense to be compatible with the measured kinematics of the MW satellites (Fig. 2 in \citealp{Boylan2012}).

Despite its theoretical appeal, the TBTF problem as stated above has its own weak points as a test of \LCDM. For example, it relies on observations of the satellites of just one galaxy, the MW. In addition, actual MW satellites are expected to be affected more by environmental effects than their DM-only counterparts. These considerations highlight the importance of assessing whether a similar issue is also present beyond the context of the MW. Recently, \citet{Tollerud2014} showed that the satellites of the Andromeda galaxy (M31) also face a TBTF problem. Based on statistical arguments, \citet{Rodriguez2013} further suggested that the TBTF problem should be generically present for MW-sized galaxies. Most importantly however, the work of \cite{Ferrero2012} showed that an analogous issue may be present for dwarf galaxies in the field\footnotemark{}. They argued that, based on the number density of galaxies measured by the Sloan Digital Sky Survey \citep[e.g.,][]{Baldry2008, LiWhite2009}, virtually all galaxies in a \LCDM \ universe should be hosted by halos with $M_{vir} \gtrsim 10^{10} \; M_\odot$. However, the rotation curves of many low-mass dwarfs indicate that they are hosted by halos below this expected mass ``threshold''. The result of \citet{Ferrero2012} is also supported by the very recent work of \citet{Kirby2014} and \citet{Garrison2014}, who show that the TBTF problem is present for non-satellite galaxies in the Local Group, as well.         

\footnotetext{In this article we use the term ``field'' loosely, to refer to galaxy samples that predominantly consist of fairly isolated galaxies. Such samples are expected to be relatively unaffected by strong environmental effects.}

In this work, we perform a similar analysis to that of \citet{Ferrero2012}, but we use a highly complementary observational dataset: the sample of galaxies detected by the Arecibo Legacy Fast ALFA (ALFALFA) survey in the emission line of atomic hydrogen (HI). This allows us to address a range of observational and theoretical uncertainties present in earlier works, which are all based on optically selected samples. For example, the current optical census of low-mass galaxies in the nearby universe is incomplete, because optical surveys are biased against low surface brightness objects. In addition, galaxies have so far been statistically connected to halos based on their measured stellar masses. However, recent observational evidence \citep{Garrison2014} suggests that this approach may not be valid in the low-mass regime.

The present article is organized as follows: in Section \ref{sec:vf+am}\footnotemark{} we describe the measurement of the galactic velocity function from the ALFALFA survey. Details regarding the measurement process can be found in Appendix \ref{sec:appendix_a}. In the same section we also describe the abundance matching (AM) procedure used for connecting galaxies to their host halos. In Section \ref{sec:dwarf_rcs} we present a sample of galaxies with resolved HI kinematics drawn from the literature, and we describe how we use their measured rotation curves to test the derived AM relation. The main result of this work is presented in \S\ref{sec:results}; readers with limited available time may choose to directly refer to this paragraph. Appendix \ref{sec:appendix_b} contains additional galactic data that are relevant for the result presented in Section \ref{sec:results}. In Section \ref{sec:discussion} we elaborate on the importance of our findings in the context of small-scale tests of \LCDM. In the same section we also point out the main uncertainties and assumptions that can have a significant impact on the results of this work. Lastly, we end with a brief summary in Section \ref{sec:summary}.

\footnotetext{The contents and layout of this section closely follow the published article of \citet{Papastergis2011}. However, the measurement of the velocity function presented here is based on a more up-to-date ALFALFA catalog \citep{Haynes2011}, and the abundance matching procedure is carried out in a more rigorous way.}

\section{Connecting observed galaxies with their host DM halos}
\label{sec:vf+am}

\subsection{Measuring the velocity function of galaxies}

\begin{figure*}
\centering
\includegraphics[scale=0.40]{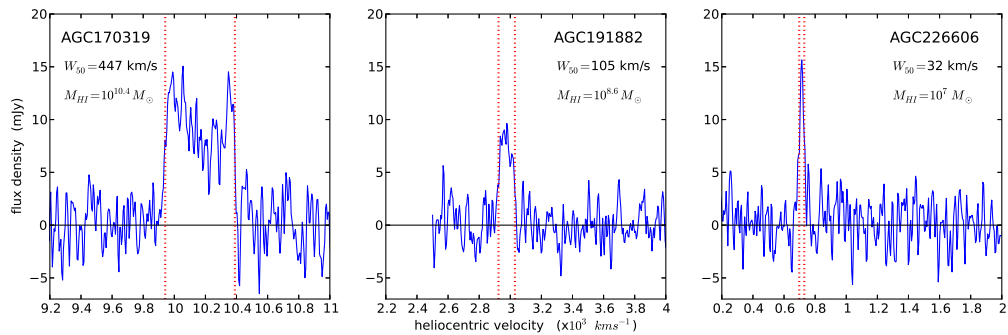}
\caption{ 
%% title: \textit{HI-line profiles for three representative ALFALFA sources.}
HI-line profiles for three representative ALFALFA sources. For each galaxy the velocity width, $W_{50}$, is measured between the two outermost points where the flux density falls to 50\% of the peak value (red vertical lines). The left panel shows a MW analog with a typical ``double horned'' profile, indicative of a mostly flat outer rotation curve. The central panel shows a dwarf galaxy with a ``boxy'' profile shape. The right panel shows an extreme dwarf galaxy with a clearly ``single-peaked'' profile, suggestive of a rising rotation curve. Keep in mind that narrow profiles can also correspond to intrinsically high-width objects that are oriented close to face-on. Note also that the velocity span of the $x$-axis in all three panels is the same (1800 \kms), and so profile widths are plotted to scale.
%HI-line profiles for three representative ALFALFA sources, spanning a large range in velocity width, $W_{50}$. The subscript $50$ refers to the fact that the width is measured between the two outermost points where the flux density falls to 50\% of the peak value (red vertical lines). The left panel shows a MW analog with a typical ``double horned'' profile, indicative of a mostly flat outer rotation curve. The central panel shows a dwarf galaxy with a ``boxy'' profile shape. The right panel shows an extreme dwarf galaxy with a clearly ``single-peaked'' profile, suggestive of a rising rotation curve. Keep in mind that narrow profiles can also correspond to intrinsically high-width objects that are oriented close to face-on. Note also that the velocity span of the $x$-axis in all three panels is the same (1800 \kms), and so profile widths are plotted to scale.
%% Conclusions: N/A
 }
\label{fig:profiles}
\end{figure*}

We measure galaxy rotational velocities using data from the publicly available catalog of the ALFALFA survey, which covers about 40\% of the final survey area \citep[\aforty \ catalog;][]{Haynes2011}. Since ALFALFA is a spectroscopic survey, each \aforty \ source has a measured spectrum of its HI line emission. Figure \ref{fig:profiles} illustrates how velocity widths, $W_{50}$, are measured from each source's lineprofile. The velocity width reflects the range of speeds at which atomic gas moves within the galactic potential (up to a projection on the line-of-sight), and therefore contains important information about the galactic kinematics.

%%%%%
%%%%%
%%%%%

\begin{figure}[htbp]
\centering
\includegraphics[width=\linewidth]{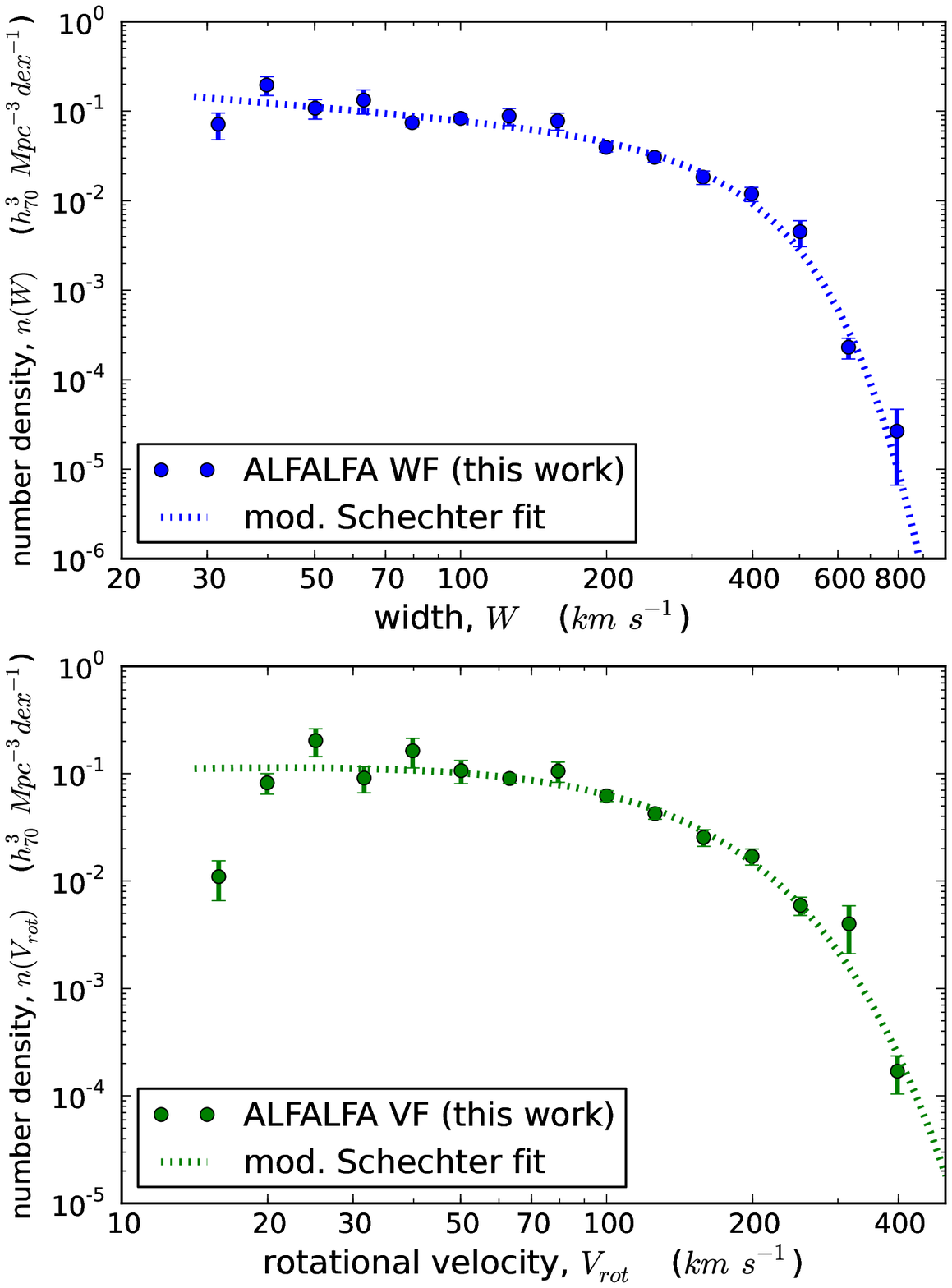}
\caption{ 
%% title: 
ALFALFA measurement of the width and rotational velocity functions.
\textit{top panel}: Data points represent the velocity width function (WF) of galaxies measured by ALFALFA, for galaxies with $M_{HI} \geqslant 10^7 \; M_\odot$. The errorbars represent the 1$\sigma$ counting noise (see Eqn. \ref{eqn:Veff}), and do not include any systematic uncertainties. The dotted line is a modified Schechter fit to the data (Eqn. \ref{eqn:mod_schechter}), with parameters: $\log(n_\ast) = -1.68 \, \pm \, 0.20 \;\; h_{70}^3 \: \mathrm{Mpc}^{-3} \mathrm{dex}^{-1}$, $\log(W_\ast) = 2.52 \, \pm \, 0.076$ \kms, $\alpha = -0.45 \, \pm \, 0.24$ and $\beta = 2.39 \, \pm \, 0.50$. \textit{bottom panel:} Data points represent the rotational velocity function (VF) of galaxies measured by ALFALFA, again for galaxies with $M_{HI} \geqslant 10^7 \; M_\odot$. Rotational velocities, $V_{rot}$, have been obtained from inclination-corrected velocity widths according to Eqn. \ref{eqn:vrot}. Errorbars again represent the 1$\sigma$ counting noise, and do not include any systematic uncertainties. The dotted line is a modified Schechter fit to the data, with parameters: $\log(n_\ast) = -1.17 \, \pm \, 0.25 \;\; h_{70}^3 \: \mathrm{Mpc}^{-3} \mathrm{dex}^{-1}$, $\log(V_\ast) = 2.04 \, \pm \, 0.18$ \kms, $\alpha = 0.14 \, \pm \, 0.41$ and $\beta = 1.46 \, \pm \, 0.37$.
%% Conclusions: N/A
 }
\label{fig:wf+vf}
\end{figure}

The top panel of Figure \ref{fig:wf+vf} shows the number density of galaxies as a function of their velocity width, as measured from the ALFALFA sample (blue datapoints). This distribution is referred to as the velocity width function (WF) of galaxies, and details on its calculation can be found in Appendix \ref{sec:appendix_a}. The measurement of the WF shown here refers to galaxies with HI masses of $M_{HI} \geq 10^7 \; M_\odot$ and velocity widths have been corrected for Doppler broadening as $W = W_{50}/(1+z_\odot)$. We then perform an analytical fit to the WF (dotted blue line), of the modified Schechter functional form:

\begin{eqnarray}
n(W) = \frac{dn_{gal}}{d\log_{10}W} = \ln(10) \; n_\ast \;  \left(\frac{W}{W_\ast}\right)^\alpha \: e^{-\left(\frac{W}{W_*}\right)^\beta} \;\; .   \label{eqn:mod_schechter}
\end{eqnarray}

\noindent 
The best fit parameters are $\log(n_\ast) = -1.68 \, \pm \, 0.20 \;\; h_{70}^3 \: \mathrm{Mpc}^{-3} \mathrm{dex}^{-1}$, $\log(W_\ast) = 2.52 \, \pm \, 0.076$ \kms, $\alpha = -0.45 \, \pm \, 0.24$ and $\beta = 2.39 \, \pm \, 0.50$. We would like to stress that the quoted errors on the parameters do not include systematic uncertainties. We would also like to note that the fit parameters are strongly covariant, such that varying each of them independently within its 1$\sigma$ range does not always produce an acceptable fit.

%% IDL mpfitfun errors
%The best fit parameters are $n_\ast = 2.07 \, \pm \, 0.62 \: \times \: 10^{-2} \;\; h_{70}^3 \: \mathrm{Mpc}^{-3} \mathrm{dex}^{-1}$, $\log(W_\ast) = 2.52 \, \pm \, 0.050$, $\alpha = -0.45 \, \pm \, 0.16$ and $\beta = 2.39 \, \pm \, 0.32$. We would like to stress that the quoted errors on the parameters do not include systematic uncertainties. We would also like to note that the fit parameters are highly covariant, such that varying each of them independently within its 1$\sigma$ range does not always produce an acceptable fit.    

The WF has the distinct advantage of measuring the distribution of velocity width, $W$, which is a direct observable for an HI-line survey. At the same time however, $W$ is measured in projection on the line-of-sight, and therefore does not correspond to any intrinsic galactic property. A more physically meaningful quantity can be derived from the inclination-corrected velocity width: if the inclination angle of a galaxy is $i$, we can define a measure of galactic rotational velocity as 

\begin{equation}
V_{rot,HI} = W / (2 \times \sin i) \; \; .    \label{eqn:vrot}
\end{equation}

Given that all ALFALFA galaxies used in this work have assigned optical counterparts, we can indeed obtain estimates of \vrot \ for each object individually. In particular, we use the SDSS counterpart's photometric axial ratio, $b/a$, to estimate galactic inclinations through the expression:

\begin{eqnarray}
\cos^2i = \frac{(b/a)^2 - q_0^2}{1-q_0^2} \;\; .
\label{eqn:incl}
\end{eqnarray}

\noindent
Here, $q_0 = 0.13$ is the assumed value of intrinsic axial ratio of galaxies viewed edge-on \citep{Giovanelli1994}. Then, the value of \vrot \ for each galaxy can be easily computed via Eqn. \ref{eqn:vrot}, and the galactic number density as a function of \vrot \ can be measured according to Eqn. \ref{eqn:Veff}. The resulting distribution is called the velocity function of galaxies (VF), and is shown in the bottom panel of Figure \ref{fig:wf+vf}. Low-inclination galaxies with $\sin i < 2/3$ were excluded from the measurement, in order to avoid making large inclination corrections. This restriction excludes about a quarter of the sample from the calculation (1657 out of 6770 galaxies), which we compensate for by uniformly increasing the normalization of the VF by the appropriate factor ($6770/5113 \approx 1.324$). We also perform a modified Schechter fit to the measured VF (green dotted line). The best fit parameters are $\log(n_\ast) = -1.17 \, \pm \, 0.25 \;\; h_{70}^3 \: \mathrm{Mpc}^{-3} \mathrm{dex}^{-1}$, $\log(V_\ast) = 2.04 \, \pm \, 0.18$ \kms, $\alpha = 0.14 \, \pm \, 0.41$ and $\beta = 1.46 \, \pm \, 0.37$. We would like to stress again that the plotted errors on the VF datapoints do not include systematic uncertainties, and therefore neither do the quoted errors on the fit parameters. Note also that we exclude from the fit the lowest velocity bin, whose very low value may be an effect of measurement incompleteness.

\subsection{Velocity abundance matching}
\label{sec:vel_am}

Both the galactic VF and WF have been extensively used in the literature to test \LCDM \ on galactic scales \citep{Obreschkow2009,Zavala2009,Zwaan2010,Papastergis2011,Trujillo2011,Obreschkow2013,Schneider2014,Klypin2014}. In particular, \LCDM \ makes a concrete prediction for the halo velocity function (HVF), i.e., the number density of halos as a function of their maximum rotational velocity, \vhalo. As long as a theoretical model exists to compute the HI rotational velocity (or velocity width) of a galaxy inhabiting a given DM halo, one can make a prediction for the galactic VF that ought to be observed. In fact, semi-analytic and semi-empirical models have been relatively successful in reproducing the observed galactic VF at intermediate and high velocities, within the context of \LCDM \ cosmology \citep{Obreschkow2009,Obreschkow2013,Trujillo2011}. However, theoretical predictions consistently overestimate the observed number density of dwarf galaxies with \vrot $\lesssim 60-80$ \kms \ \citep{Zavala2009,Trujillo2011,Schneider2014}. This is true even for more refined models based on hydrodynamic simulations, which have trouble reproducing observations at low velocities as well \citep[see, e.g.,][Fig. 10]{Sawala2013}. At the same time, it is important to keep in mind that the models mentioned above implicitly assume that observed values of \vrot \ are effectively measuring the maximum rotational velocity of the host halos.

\begin{figure}
\centering
\includegraphics[scale=0.42]{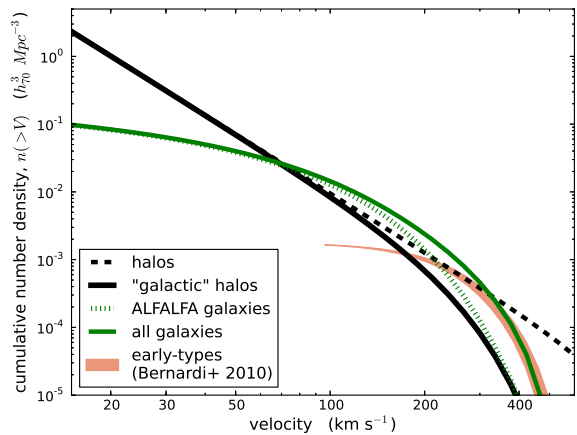}
\caption{ 
%% title: 
Cumulative velocity functions for galaxies and \LCDM \ halos.
The dashed black line represents the cumulative number density of halos as a function of their maximum circular velocity, \vhalo, in a \LCDM \ universe (BolshoiP simulation; \citealp{Hess2014}). The solid black line corresponds to the ``galactic'' HVF, which excludes massive halos that are unlikely to host individual galaxies (see Eqn. \ref{eqn:gal}). The dotted green line shows the cumulative VF of ALFALFA galaxies. The red shaded region represents the VF of early-type galaxies, measured from SDSS data \citep{Bernardi2010}. The solid green line represents the VF of galaxies of all morphological types, obtained from the sum of the ALFALFA and early-type VFs.
%% Conclusions: N/A
}
\label{fig:cum_vfs}
\end{figure}

In view of the difficulties faced by theoretical models in reproducing the observed VF of galaxies, we take here a more conservative approach: we use the galactic VF measured by ALFALFA to infer the expected relation between \vrot \ and \vhalo \ in a \LCDM \ universe. Our approach therefore does away with the assumption that \vrot \ is approximately equal to \vhalo. 

We then go ahead and infer a quantitative \vrotvhalo \ relation by using the statistical technique of abundance matching (AM). AM operates under the assumption that galaxies with higher \vrot \ are hosted on average by halos with higher \vhalo \ (see discussion in \S\ref{sec:stochasticity}). Thus $V_h(V_{rot})$ can be derived by matching the cumulative number densities of galaxies and halos, i.e., by demanding that

\begin{equation}
n_{gal}(>V_{rot}) = n_h(>V_h(V_{rot})) \;\; . 
\label{eqn:am}
\end{equation}

\noindent
The number density of halos on the right hand side of the equation above also includes the contribution of subhalos.

Figure \ref{fig:cum_vfs} shows the cumulative velocity distributions of galaxies and halos used in the AM procedure. Specifically, we use the halo velocity function, $n_h(>V_h)$, of the BolshoiP \LCDM \ simulation \citep{Hess2014}. The BolshoiP simulation assumes a \textit{Planck} first-year cosmology \citep{Planck2013}, and therefore predicts a higher normalization for the halo velocity function (HVF) than its WMAP7 predecessor \citep[Bolshoi simulation;][]{Klypin2011}. The cumulative VF of ALFALFA galaxies is also shown in Fig. \ref{fig:cum_vfs}. It is calculated by integrating the fit to the differential galaxy VF (see bottom panel of Fig. \ref{fig:wf+vf}). It is important to notice that the HVF in a \LCDM \ universe is significantly steeper than the observed galactic VF, which is the reason why theoretical models generically overpredict the number of low-velocity galaxies.

Before we can proceed, we need to address here two complications: 
Firstly, massive halos with $M_{vir} \gtrsim 2\times10^{13} \; M_\odot$ are usually not the hosts of individual galaxies, but rather of galactic groups or clusters. We therefore use the factor 

\begin{equation}
f_g(V_h) = e^{-(V_h/330 \: \mathrm{km}\mathrm{s}^{-1})^3} \;\; 
\label{eqn:gal}
\end{equation}

\noindent
to suppress the high-velocity end of the HVF, $n_g(V_h) = f_g(V_h) \times n(V_h)$.
The resulting ``galactic'' HVF approximately matches the results of \citet{Shankar2006} and \citet{Rodriguez2011}. 

%We therefore calculate a differential ``galactic'' HVF as $n_g(V_h) = f_g(V_h) \times n(V_h)$, where: 

%\begin{equation}
%f_g(V_h) = e^{-(V_h/330 \: \mathrm{km}\mathrm{s}^{-1})^3} \;\; .
%\label{eqn:gal}
%\end{equation}

%\noindent
%The factor above is used to suppress the high-velocity end of the HVF, and provides an approximate match with the results of \citet{Shankar2006} and \citet{Rodriguez2011}. The cumulative version of the ``galactic'' HVF is shown as a thick black solid line in Fig. \ref{fig:cum_vfs}.   

The second complication stems from the fact that blindly selected HI samples have a bias against gas-poor, early-type galaxies. This can be seen directly by placing ALFALFA detections on a color-magnitude diagram \citep[Fig. 10]{Huang2012b}, but also indirectly by measuring the clustering properties of ALFALFA galaxies \citep[Fig. 13]{Papastergis2013}. To compensate for this effect we consider the work of \citet{Bernardi2010}, who have used SDSS spectra to measure stellar velocity dispersions for galaxies of different morphological types. Their results for the cumulative VF for early-type galaxies is shown in Fig. \ref{fig:cum_vfs}. We use two empirical relations to transform the SDSS-based velocity dispersions into equivalent rotational velocities: Eqn. 3 in \citet{Baes2003} and Eqn. 2 in \citet{Ferrarese2002} (with the parameters appropriate for elliptical galaxies as given in their Table 2). The plotted range of the early-type VF reflects the difference between the two relations, and illustrates the typical (small) uncertainty associated with such transformations.

Lastly, we obtain the VF for all galaxies (regardless of type) from the sum of the ALFALFA VF and the VF of early-type objects measured by Bernardi et al. Keep in mind that the measurement of Bernardi et al. is restricted to velocities $\gtrsim 100$ \kms; Therefore, we implicitly assume that the HI-selected ALFALFA sample is a complete census of low-velocity galaxies (but see \S\ref{sec:early_types} for a discussion).

\begin{figure}[htbp]
\centering
\includegraphics[width=\linewidth]{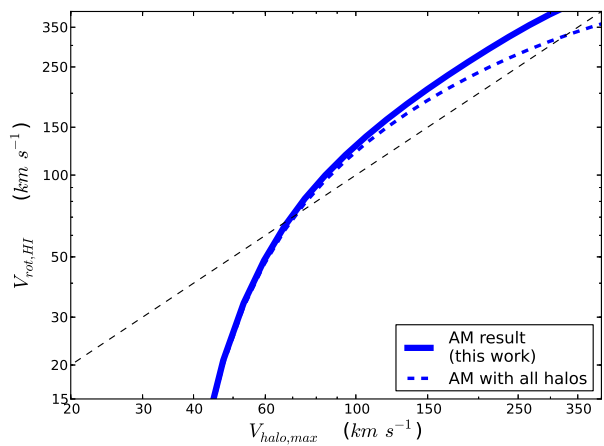}
\caption{
%% title
\vrotvhalo \ relation in a \LCDM \ universe.
The blue line is the average relation between the galactic HI rotational velocity (\vrot) and the maximum circular velocity of the host halo (\vhalo), in a \LCDM \ universe. This relation is obtained by abundance matching (AM), i.e., by matching the cumulative velocity distributions of all galaxies and ``galactic'' halos (see Fig. \ref{fig:cum_vfs}). The dashed blue line is the AM result when galaxies are matched to all halos. The dashed black line is a reference one-to-one line.
%% Conclusions:
%Note the marked ``downturn'' of the relation at low velocities: it is needed to reconcile the steeply rising HVF predicted by \LCDM \ with the much shallower galactic VF measured by ALFALFA. 
}
\label{fig:vrotvhalo}
\end{figure}

In Figure \ref{fig:vrotvhalo} we present the average \vrotvhalo \ relation expected in a \LCDM \ universe, derived through AM. Specifically, the relation is obtained by matching the cumulative number densities of ``galactic'' halos and galaxies of all types (see Fig. \ref{fig:cum_vfs}). \vrot \ is greater than \vhalo at intermediate and high velocities, reaching maximum ratios of $V_{rot}/V_h \approx 1.4$. Ratios larger than unity can be achieved when there is a large baryonic contribution to the galactic rotation curve (RC), which leads to a substantial boost to the rotational velocity \citep[see e.g.,][Fig. 5]{Trujillo2011}. Keep in mind however, that the exact behavior of the AM relation at velocities \vrot$\gtrsim 250$ \kms \ is relatively uncertain. For example, Fig. \ref{fig:vrotvhalo} also shows the AM result when all halos are included. Moreover, the addition of a modest amount of scatter to the idealized Eqn. \ref{eqn:am} will have a similar effect, namely it will alter the high-velocity end of the relation.

Most importantly however, the relation displays a remarkable ``downturn'' at low velocities, whereby the \vrot \ of dwarf galaxies is predicted to significantly underestimate the \vhalo \ of their hosts. This effect is expected to become more pronounced as the measured HI velocity decreases: for example, galaxies with \vrot$= 15 - 20$ \kms \ are expected to reside in halos with \vhalo $\approx 40 - 45$ \kms. This behavior has a simple and intuitive explanation: if halos with \vhalo $< 40$ \kms \ were allowed to host galaxies that are detectable by ALFALFA, then the galactic number density in a \LCDM \ universe would be much larger than observed (refer to Fig. \ref{fig:cum_vfs}). The discussion above does not change even in the presence of scatter in the \vrotvhalo \ relation, because scatter does not affect the AM result at the low end\footnotemark{}.

\footnotetext{Including scatter in Eqn. \ref{eqn:am} is equivalent with deconvolving the VF of Fig. \ref{fig:wf+vf} with a lognormal distribution. This process will strongly affect the high-velocity end of the VF (which is an exponential fall off), but will minimally affect the low-velocity end (which is a power law).}

\section{Internal kinematics of dwarf galaxies}
\label{sec:dwarf_rcs}

\subsection{Data sample}
\label{sec:data_sample}

In this section we aim to place individual galaxies on the \vrotvhalo \ diagram shown in Fig. \ref{fig:vrotvhalo}, in order to observationally test the relation predicted by \LCDM. Ideally, one would like to place on the diagram the same objects that went into calculating the galactic VF; this would ensure that the AM relation is derived using the same objects that are used for testing it. Unfortunately however, the data provided by ALFALFA are not sufficient to accomplish this task. In particular, even though ALFALFA data can be used to derive values of \vrot \ (in conjunction with optical inclinations), they do not contain enough information to constrain \vhalo.

We therefore use instead an extensive sample of galaxies with interferometric HI observations, gathered from the literature. In particular, we use 12 galaxies from \citet{Sanders1996}, 30 galaxies from \citet{VerheijenSancisi2001}, 19 galaxies from the The HI Nearby Galaxy Survey (THINGS; \citealp{deBlok2008, Oh2011}), 54 galaxies from the Westerbork HI Survey of Spiral and Irregular Galaxies (WHISP; \citealp{Swaters2009, Swaters2011}), 12 galaxies from the Local Volume HI Survey (LVHIS; \citealp{Kirby2012}), 5 galaxies from \citet{Trachternach2009}, 4 galaxies from \citet{Cote2000}, 28 galaxies from the Faint Irregular Galaxies GMRT Survey (FIGGS; \citealp{Begum2008a,Begum2008b}), 17 galaxies from the LITTLE THINGS survey (\citealp{Hunter2012}; Oh et al., in prep.), 11 galaxies from the SHIELD survey \citep{Cannon2011}, and the recently discovered galaxy Leo P \citep{Bernstein2014}.

The sample of individual objects assembled above is inhomogeneous, but spans an impressive range of observed rotational velocities: from $> 300$ \kms \ for galaxy UGC2885 in the \citet{Sanders1996} sample to $\approx 12$ \kms \ for galaxy KK44 in the FIGGS sample. This means that the kinematics of galaxies of very different sizes are modeled to different levels of detail. For example, most of the spiral galaxies in the THINGS sample analyzed by \citet{deBlok2008} have RCs of exquisite resolution, and mass models constrained by multiwavelength data. On the other hand, the velocity fields of the extremely low-mass SHIELD galaxies are significantly harder to model, and only outermost-measured-point velocity estimates are available for them at present. We would like to note that the objects in the literature sample do not share the same selection rules as the \aforty \ sample; for example, some of the galaxies with interferometric observations have $D < 7$ Mpc or $M_{HI} < 10^7 \; M_\odot$. Keep also in mind that all galaxies have been observed as part of targeted HI interferometric campaigns. As a result, almost all galaxies are relatively gas-rich and fairly isolated; this means that the vast majority of the objects are field galaxies.

\subsection{Kinematic analysis}
\label{sec:kinematic_analysis}

\begin{figure}[htbp]
\centering
\includegraphics[width=\linewidth]{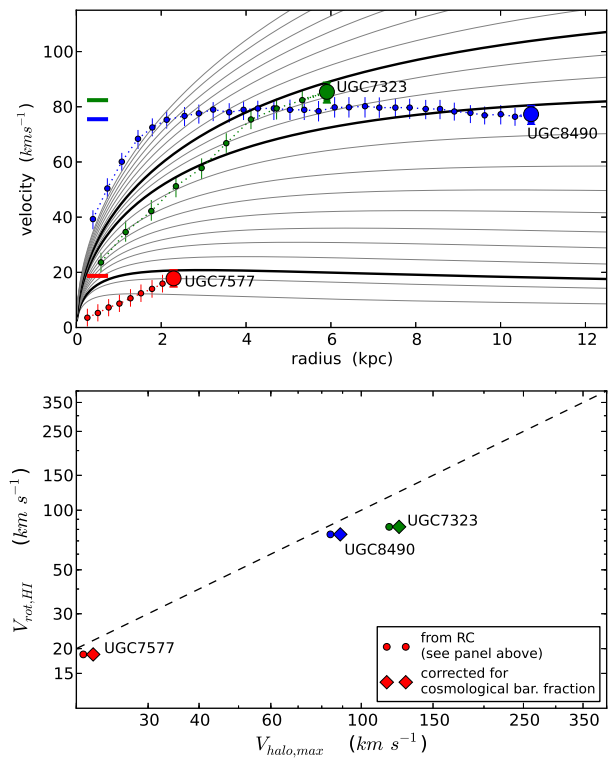}
\caption{ 
%% title:
Kinematic analysis of galaxies with HI rotation curves.
\textit{top panel:} Three representative rotation curves (RCs) from the literature sample of galaxies with interferometric HI observations (red: UGC7577, green: UGC7323; blue: UGC8490). The gray solid lines show the RCs of increasingly more massive NFW halos (from bottom to top). The concentration parameter for each halo is the median concentration expected for its mass in a \textit{Planck} cosmology \citep{DuttonMaccio2014}. The three thick gray lines show the most massive NFW halos that are compatible with the last measured point of each galaxy's RC (large colored symbols), to within 1$\sigma$. For each galaxy, only the last measured point of the RC is used for the analysis in this work. The peak velocity for each of these three halos is the value of \vhalo \ assigned to the corresponding galaxy. The colored horizontal marks denote the value of \vrot \ for the three galaxies, inferred from their inclination-corrected linewidths (see Eqn. \ref{eqn:vrot}). \textit{bottom panel:} Small circles show the placement of UGC7577, UGC7323 and UGC8490 on the \vrotvhalo \ diagram, according to the procedure outlined in the top panel. The large diamonds show their positions after a correction is applied to account for the cosmic baryon fraction (see text for details).
%%Conclusions: Note that galaxies UGC7323 and UGC8490 have very similar values of \vrot, but significantly different values of \vhalo; as the upper panel shows, the extended rotation curve of UGC8490 can place a more stringent constraint on \vhalo \ than the less spatially extended RC of UGC7323. 
 }
\label{fig:find_vhalo}
\end{figure}

Figure \ref{fig:find_vhalo} illustrates the process used to place galaxies on the \vrotvhalo \ diagram. We use three galaxies as examples, taken from the \citet{Swaters2011} sample: UGC7577, UGC7323 and UGC8490. Their RCs are shown in the top panel of Fig. \ref{fig:find_vhalo}. For each galaxy, the last measured point (LMP) of its RC is marked with a bold symbol; this is the \textit{only} datapoint used in this work to constrain \vhalo.

We then consider a set of DM halos with monotonically increasing masses. The DM halos are assumed to have an NFW mass profile \citep{NFW1997} of the form

\begin{equation}
\rho(R) = \frac{\rho_0}{\frac{R}{R_s}\left[1+\left(\frac{R}{R_s}\right)^2\right]} \;\; .
\end{equation}

\noindent
$R_s$ and $\rho_0$ are the scale radius and scale density, respectively. In practice, halos are more often described in terms of virial quantities: in this work we define the virial radius, $R_{vir}$, at a density contrast of 104 with respect to the critical density\footnotemark{} ($R_{vir} = R_{104c}$). Then we can define the virial velocity as the circular velocity of the halo at the virial radius, $V_{vir} = V(R_{vir})$, and the concentration parameter as $c=R_{vir}/R_s$. Therefore $V_{vir}$ and $c$ describe ``how massive'' and ``how dense'' a halo is, respectively. The RC of an NFW halo can now be written as

\footnotetext{This is the density contrast value corresponding to a \textit{Planck} cosmology at  $z=0$, according to the spherical collapse model \citep{Mainini2003}. Other definitions of the virial radius are also commonly used, for example referring to a density contrast of 200 with respect to the critical density ($R_{200c}$) or with respect to the cosmic matter density ($R_{200m}$). We would like to note that the analysis in this work does not depend on which virial definition is adopted. This is because all relevant quantities related to DM halos, namely \vhalo \ and the halo RC, are physical and therefore independent of the virial definition. Virial quantities are merely a convenient way to parametrize the halo RC and the mass-concentration relation.}

\begin{equation}
V_c(R) = V_{vir} \times \sqrt{\frac{\ln(1+cx) - \frac{cx}{1+cx}}{x\left(\ln c - \frac{c}{1+c}\right)}} \;\; ,
\label{eqn:nfw_rc}
\end{equation}

\noindent
where $x=R/R_{vir}$ is the normalized radius. 

We identify the most massive halo that is compatible with the velocity measured at the LMP for each of the three example galaxies, to within the 1$\sigma$ velocity error. 
The peak RC amplitude for each of these three halos is the value of \vhalo \ assigned to the corresponding galaxy. It is important to keep in mind that for most objects the maximum halo velocity is reached beyond the extent of the galaxy's RC. For example, the RC of UGC7323 extends to $R=5.9$ kpc, but its host halo achieves its maximum velocity of $117.1$ \kms \ at $R = 32.9$ kpc.  
The procedure above does not assume a priori that \vrot$\approx$\vhalo, as is usual practice in many theoretical works. In fact, galaxies with less extended RCs (e.g., UGC7323), can be assigned \vhalo \ values that are significantly higher than their \vrot.
%% INFO:
% U7577:  Vh,max = 20.8  Rmax = 2.93
% U8490:  Vh,max = 84.0  Rmax = 20.73
% U7377:  Vh,max = 117.1  Rmax = 32.94

The values of \vrot \ for each example galaxy are computed based on the observed width of the galactic lineprofile, $W$, and the galaxy inclination, $i$ (Eqn. \ref{eqn:vrot}). The profile widths of galaxies in our literature sample can be obtained from either their interferometric or previous single-dish observations. Their inclinations, $i$, can be derived in a variety of ways, depending on the data quality: When the galactic velocity field is well resolved, the inclination can be measured by modeling the observed kinematics. When this is not possible, one can use the axial ratio of the spatially resolved HI emission to infer the inclination of the gaseous disk. If neither method is practical, the photometric axial ratio of the optical counterpart can be used instead (as per Eqn. \ref{eqn:incl}). In general, kinematic inclinations and inclinations based on the HI axial ratio are thought to be more accurate than optical ones \citep{Kirby2012,VerheijenSancisi2001}.

The bottom panel of Fig. \ref{fig:find_vhalo} shows the positions of the three example galaxies on the \vrotvhalo \ diagram, based on the analysis described above. Note that for a fixed value of \vrot, galaxies with a more extended RC can place a more stringent constraint on \vhalo. This is immediately obvious by comparing the position of galaxies UGC8490 and UGC7323 on the \vrotvhalo \ diagram, and noting the extent of their RCs in the top panel of Fig. \ref{fig:find_vhalo}.

Let us now point out two complications related to the procedure described above: First, the AM result refers to \vhalo \ values for the halos in the BolshoiP simulation, where the total matter content of the universe ($\Omega_m = \Omega_{DM} + \Omega_{bar}$) is treated as a dissipationless fluid. This means that, if $f_{bar} \approx 0.15$ is the cosmic baryon fraction, a realistic halo would have a factor of $(1-f_{bar})$ less DM mass than its corresponding BolshoiP halo. The rest of the mass would be in the form of baryons, part of which will end up in the galactic disk in the form of stars or gas. Ideally, one would subtract the contribution of baryonic components from a galaxy's RC before using it to constrain the host halo mass. However, here we do not attempt to calculate baryon-subtracted RCs; we make instead the conservative assumption that the RCs of all galaxies in our sample are fully attributable to DM. Therefore, we simply increase the derived value of \vhalo \ by a factor of $(1-f_{bar})^{-1/3}$, in order to reflect the higher DM content of the simulation (the $1/3$ exponent follows from the $M_{vir} \propto V_h^3$ scaling for DM halos). 
%The new positions of the three example galaxies in the \vrotvhalo \ diagram are marked in the bottom panel of Fig. \ref{fig:find_vhalo} with large diamonds. 
These new \vhalo \ estimates are overly conservative in the case of large spiral galaxies, whose RCs have a sizable contribution from baryons. However, the rescaled \vhalo \ values are fairly realistic for dwarf galaxies, which are typically DM-dominated \citep[e.g.,][]{Papastergis2012,Sawala2013}.

Second, as one can realize from inspection of Eqn. \ref{eqn:nfw_rc}, it is not possible to determine a unique value for \vhalo \ given only one point of the galactic RC. The reason is that the RCs of NFW halos are a two-parameter family: one can find a wide range of compatible halos, by simply adjusting the values of $V_{vir}$ and $c$. In this work, we fix $c$ to the median value as a function of halo mass, as expected in a \textit{Planck} cosmology \citep[Table 3 in][]{DuttonMaccio2014}.

In principle, one could use the whole RC of a galaxy to simultaneously fit for \vhalo \ and $c$. In fact, such an analysis is available for several galaxies in our sample \citep{deBlok2008,Oh2011,Swaters2011}. However, as shown by the RC of UGC7577, these studies find that NFW halos of median concentration do not provide a good description of the inner velocity profiles of dwarf galaxies \citep[e.g.,][]{Oh2011}. Recently, hydrodynamic simulations have shown that the inner DM profiles of halos can be affected by baryonic feedback processes \citep[e.g.,][]{Governato2010}, resulting in RCs that better match the observations of the inner velocity profiles of dwarfs \citep{Oh2011b}. Consequently, one may repeat the kinematic analysis described in this section using instead a velocity profile that captures the modification of the galactic RC induced by feedback \citep[e.g.,][]{diCintio2014a,diCintio2014b}. In such case, the value of \vhalo \ assigned to a given galaxy can, in principle, be substantially different from the value obtained from an NFW analysis \citep{BrookdiCintio2014}. This fact represents an important caveat in the kinematic analysis described in this section, which we would like to fully acknowledge. At the same time, the NFW analysis can still produce reliable results, provided that the difference between the modified profile and its unmodified NFW counterpart is small at the radius of the galaxy's LMP (see Appendix \ref{sec:appendix_b}). We devote \S\ref{sec:baryon_rc} to assess to what extent the effect of feedback on the velocity profiles of halos can affect the results of this article.

\subsection{Results}
\label{sec:results}

\begin{figure*}[htbp]
\centering
\includegraphics[scale=0.49]{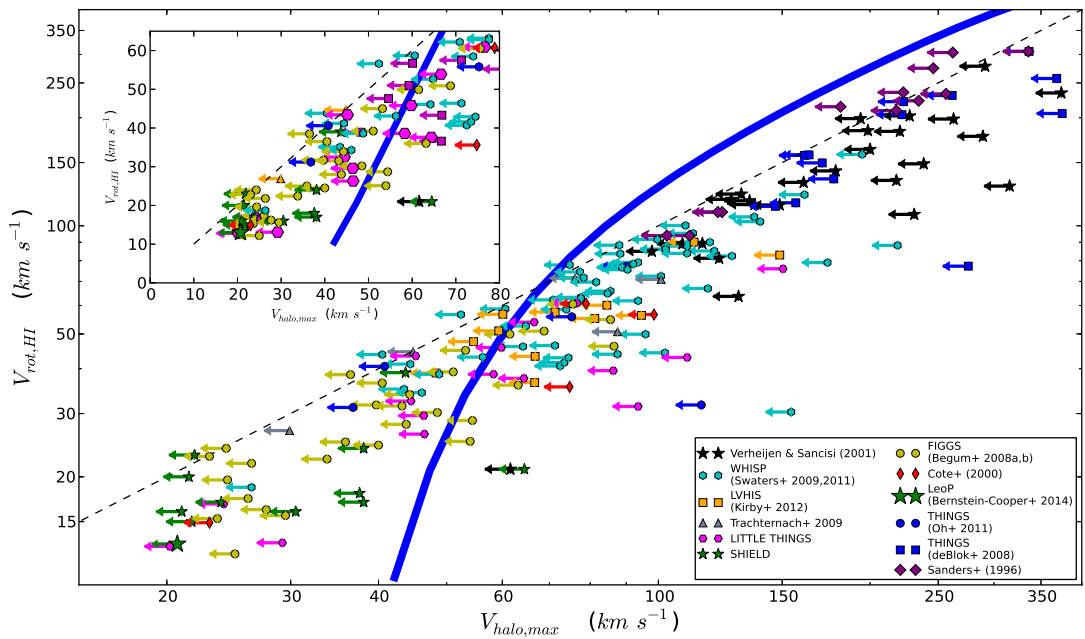}
\caption{
%% title: 
Placement of galaxies on the \vrotvhalo \ diagram.
\textit{main figure:} The blue line is the average \vrotvhalo \ relation in a \LCDM \ universe, inferred from abundance matching (same as in Fig. \ref{fig:vrotvhalo}). The colored points represent a sample of 194 galaxies with interferometric HI observations, drawn from the literature. Their \vrot \ and \vhalo \ values are computed as described in Fig. \ref{fig:find_vhalo}. All points are drawn as upper limits, because we make the conservative assumption that the contribution of baryons to the galactic RC is negligible for all galaxies. 
 %% Conclusions: Note that at very low velocities (\vrot $< 25$ \kms) \emph{all but three} of the upper limits are inconsistent with the AM relation. This situation is analogous to the ``too big to fail'' problem faced by the bright satellites of the MW, but here is seen for galaxies in the field. 
Refer to \S\ref{sec:results} and Sec. \ref{sec:discussion} for the scientific interpretation of this figure.
\textit{inset panel:} A zoom-in on the low-velocity region of the diagram (linear axes). }
\label{fig:vrotvhalo+gals}
\end{figure*}

Figure \ref{fig:vrotvhalo+gals} shows the placement on the \vrotvhalo \ diagram of all the 194 galaxies with literature HI rotation curves. Twelve galaxies have duplicate entries in more than one of the samples listed in \S\ref{sec:data_sample}, in which cases we plot the observation corresponding to the most extended RC (taking data quality into consideration as well). All galaxies are positioned as upper limits on the diagram, because their \vhalo \ values refer to the most massive compatible halo. 

Over most of the range in HI rotational velocity, 30 \kms $\lesssim$ \vrot $\lesssim$ 300 \kms, the AM relation is consistent with the upper limits obtained from individual galaxies. Most importantly though, the situation changes at the lowest velocities probed: For example, when one considers galaxies with \vrot $< 25$ \kms, \emph{all but three} of the upper limits are inconsistent with the AM relation. We have verified that this result holds even if we substitute median concentration halos in our kinematic analysis with 2$\sigma$ under-concentrated ones. The inconsistency arises because the AM relation predicts fairly massive hosts for the lowest-velocity galaxies in our sample; however, such massive halos would exceed the velocity measured at the LMP for these objects. A different way of phrasing the inconsistency is that, if halos with \vhalo$\lesssim 30$ \kms \ were allowed to host the lowest-velocity galaxies in our sample (as the galactic kinematics indicate), then their number density in a \LCDM \ universe would be much higher than what observed by ALFALFA. Fig. \ref{fig:vout_radii} offers yet another way to visualize the inconsistency in an intuitive fashion.

Before we proceed with a discussion of the scientific relevance of this result, we would like to point out a number of subtleties related to the positioning of objects on the \vrotvhalo \ diagram:

First, the RCs of DM halos described by Eqn. \ref{eqn:nfw_rc} represent the halo circular velocity at each radius, which reflects the enclosed dynamical mass. However, the gas in actual galaxies undergoes some turbulent motion in addition to rotation, with typical amplitudes of $8-10$ \kms. As a result, observed RCs should be corrected for pressure support before being compared to theoretical halo RCs. These corrections are most important for the lowest velocity galaxies in our sample, where ordered rotation and turbulent motion have similar amplitudes. Pressure support corrections have been performed in the original sources for the galaxies in the THINGS, LITTLE THINGS, FIGGS and \citet{Cote2000} samples, as well as for LeoP. On the other hand, no corrections have been originally applied to the galaxies in the WHISP and SHIELD samples. For galaxies in these two samples we apply a crude pressure support correction to their LMP, $V_{\mathrm{LMP}} \rightarrow \sqrt{V_{\mathrm{LMP}}^2 + 2  \sigma^2}$, assuming $\sigma = 8$ \kms. The correction increases the LMP velocity somewhat, and therefore results in a slightly higher value of \vhalo.

Second, galactic RCs have been checked for their quality, to the extent allowed by the material published in each original reference. In cases where the velocity for the originally published outermost radius was deemed unreliable, we adopted a measurement at a smaller radius as our LMP. Even though this process is highly subjective, any truncation of the original RC is conservative for the purposes of this work (see \S\ref{sec:kinematic_analysis}). Note also that for the extreme dwarf galaxies of the FIGGS, LITTLE THINGS and SHIELD samples no quality control was performed, since the only available data were LMP radii and velocities.

Third, we have given no information regarding the observational errors associated with the placement of each galaxy on the \vrotvhalo \ diagram. Unfortunately, we cannot quantify the errors on both \vrot \ and \vhalo \ for each galaxy in a rigorous way, but we would like to offer some qualitative guidance. The error on \vrot \ is determined by the measurement error on the velocity width and the uncertainty on the adopted inclination value. These errors are not always quoted in the original references. Typical values for the former are a few \kms, while values for the latter depend on the method used to determine the inclination ($\approx 5^\circ$ for kinematic inclinations of good quality, larger in other cases). In the case of low inclination galaxies, a small error in inclination can translate into a fairly large error on \vrot. Keep also in mind that it is difficult to quantify the systematic component of the inclination uncertainty, especially when different methods are used for different objects. 

Errors on \vhalo \ are even harder to quantify: They depend on the error with which the LMP velocity and radius are measured. The LMP velocity error mainly depends on the inclination uncertainty (discussed above). However, keep in mind that if the velocity field is asymmetric, or if there are significant non-circular motions, systematic uncertainties arise that are difficult to quantify. The LMP radius, on the other hand, is affected by distance uncertainties; this is because the conversion from an angular extent on the sky to physical units (kpc) is distance-dependent. The uncertainty on the distance varies a lot form object to object: some galaxies in our sample have accurate primary distances (e.g., TRGB), while others have lower accuracy distances based on flow models or even pure Hubble flow. 

In general, errors on inclination affect both \vrot \ and \vhalo. As a result, galaxies move roughly diagonally on the \vrotvhalo \ diagram. On the other hand, errors on the distance only affect \vhalo. This causes galaxies to move horizontally on the diagram.

\section{Discussion}
\label{sec:discussion}

The TBTF issue was first identified in the MW system, as an incompatibility between the measured kinematics of the brightest MW satellites and the kinematics of their expected host subhalos in \LCDM \ simulations \citep{Boylan2011}. However, a number of possible solutions to the MW TBTF problem within the \LCDM \ model have been identified, thus disputing the cosmological significance of the discrepancy. For example, \citet{Wang2012} and \citet{Vera2013} have argued that if the mass of the MW halo is $M_{vir} < 1 \times 10^{12} \, M_\odot$ then the TBTF problem would likely not occur. This is because the typical masses of the largest subhalos scale sensitively with the mass of the host halo. A MW mass in this range is on the low side of observational estimates \citep[e.g.,][]{Watkins2010}, and is lower than typically assumed in DM simulations of MW analogs. Another solution can come from considering the cosmic variance associated with observations of a single object. \citet{PurcellZentner2012} have argued that the TBTF problem is expected to occur in at least 10\% of MW-sized halos just due to halo-to-halo variation in the subhalo population. 

The plausibility of the two solutions above has since been put into question, because the TBTF problem is likely present in the satellite populations of galaxies other than the MW. For example, \citet{Tollerud2014} finds that the TBTF problem is also present in the satellite system of the Andromeda galaxy (M31). This finding weakens the ``light'' MW argument, because it is unlikely that both the MW and Andromeda are hosted by halos with $M_{vir} < 1 \times 10^{12} \; M_\odot$ \citep{vdMarel2012}. It also weakens the cosmic variance argument, because it is improbable that both the MW and M31 are outliers in terms of their subhalo populations. In addition, \cite{Rodriguez2013} find based on statistical considerations that the MW satellites are a fairly typical population for a galaxy of this size \citep[see also][]{StrigariWechsler2012}.

Nonetheless, a different potential solution to the TBTF has been put forward by \citet{Zolotov2012}, which is generically applicable to the satellite population of any MW-sized halo. This solution is related to baryonic effects that had not been taken into account in the original TBTF formulation. In particular, Zolotov et al. argue that internal feedback processes in low-mass halos (e.g., gas blowout due to star formation) will lead to the formation of low-density ``cores'' in their inner DM profiles. This fact, in conjunction with the presence of a stellar disk in the MW, will lead to significantly enhanced tidal stripping of subhalos compared to the DM-only case. As a result, a significant amount of mass can be removed from the central parts of subhalos, leading to velocity profiles that are consistent with measurements (see Fig. 3 in \citealp{BrooksZolotov2014}). This baryonic solution to the TBTF problem has been regarded as a generic and robust way to resolve the discrepancy. However, the proposed mechanism relies on processes that are specific to satellite galaxies; this is why establishing whether the TBTF problem is also present for field galaxies has important scientific implications.

The first evidence for a positive answer came from the work of \citet{Ferrero2012}. In particular, they used the stellar mass function (SMF) of galaxies to infer an $M_\ast$-$M_h$ relation in a \LCDM \ universe, via the technique of abundance matching. They then showed that the rotation curves of gas-rich galaxies with low stellar masses ($M_\ast \lesssim 10^7 \; M_\odot$) cannot accommodate host halos as massive as expected in \LCDM \ (see their Fig. 3). The present work confirms the results of Ferrero et al., and at the same time addresses a number of systematic uncertainties present in their analysis. First, the SMF measured by current wide-area optical surveys, such as the SDSS, suffers from surface brightness incompleteness at low stellar masses ($M_\ast \lesssim 3 \times 10^8 \; M_\odot$; see Fig. 6 in \citealp{Baldry2008}). As a result, the low-mass end of the SMF could be much steeper than the measured one, in which case the discrepancy reported by Ferrero et al. could be significantly alleviated or perhaps resolved \citep[see][Fig. 4]{Ferrero2012}. In the contrary, the measurement of the galactic VF from the ALFALFA HI-selected sample does not share the same surface brightness incompleteness; in fact, low surface brightness galaxies in the field are expected to be gas-rich, and therefore easily detectable by ALFALFA. Barring therefore the presence of a dominant population of low surface brightness \textit{and} gas-poor galaxies in the field, the present estimates for the number density of low-mass galaxies seem robust (see also \S\ref{sec:early_types}).

In addition, the AM procedure in Ferrero et al. is performed on the basis of stellar mass, as is the case in all previous works assessing the TBTF problem \citep{Boylan2011,Tollerud2014,Garrison2014}. This implicitly assumes that there is a monotonic relation between $M_\ast$ and $M_h$ (or \vhalo). However, \citet{Garrison2014} have used the stellar kinematics of local group galaxies to investigate whether such a relation is indeed present at very low masses; under the assumption of a universal halo mass profile, they find that for galaxies with $M_\ast \lesssim 10^{8} \; M_\odot$ the stellar mass shows very little correlation with the inferred \Vhalo \ (see their Fig. 12). By contrast, Fig. \ref{fig:vrotvhalo+gals} seems to justify the velocity-based AM procedure used in this work. In particular, the datapoints show a clear monotonic trend between \vrot \ and \vhalo, down to the lowest velocities probed. In addition, \vrot \ values at fixed \vhalo \ show a well-behaved and relatively small scatter\footnotemark{} of 0.1 dex.

\footnotetext{The scatter value mentioned here is calculated by eye, as 1/4 of the \vrot \ range encompassing most upper limits at fixed \vhalo \ (excluding outliers).}

Lastly, most of the extremely low-velocity HI rotation curves analyzed by Ferrero et al. belong to galaxies in the FIGGS sample. This means that their results regarding the low-velocity end are susceptible to systematics or selection effects that are specific to this one sample. In this work we have significantly increased the number of extreme dwarf galaxies, by adding objects from the SHIELD and LITTLE THINGS projects. Reassuringly, Figure \ref{fig:vrotvhalo+gals} shows no discernible differences among the various dwarf samples.

Apart from the observational factors mentioned in \S\ref{sec:results}, a number of other uncertainties and assumptions may affect the analysis performed this work. In the remainder of this section we consider in detail several such issues, and we show that they do not have a large impact on our main conclusions (at least when considered individually).

\subsection{Measurement uncertainties on the galactic VF?}
\label{sec:uncertain_vf}

An accurate determination of the galactic VF is of great importance in the context of Fig. \ref{fig:vrotvhalo+gals}: in particular, the measured number density of galaxies with low velocities determines the exact behavior of the \vrotvhalo \ AM relation at the low-velocity end.

\begin{figure}
\centering
\includegraphics[scale=0.51]{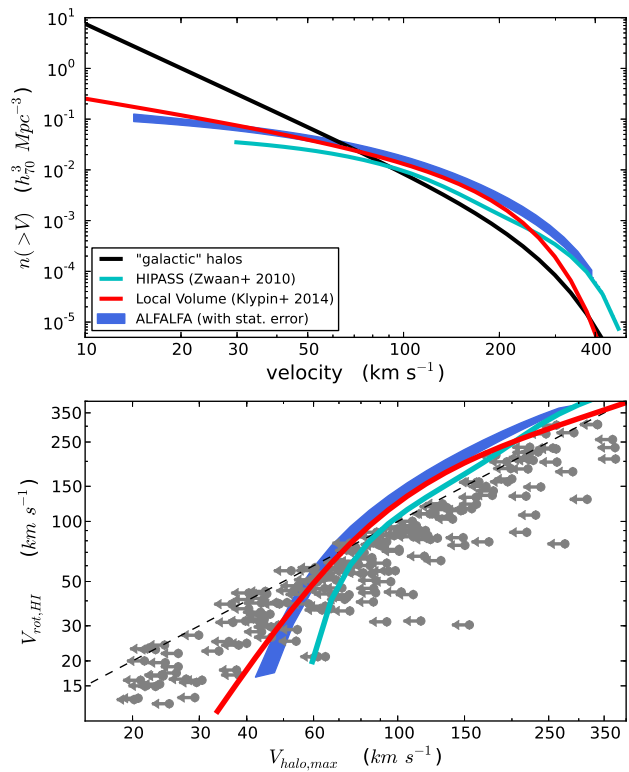}
%% title: 
\caption{ Observational uncertainties on the VF measurement. \textit{top panel:} The blue shaded region shows the range of ALFALFA cumulative VFs that correspond to modified Schechter parameters within 1$\sigma$ from the best fit values (see Eqn. \ref{eqn:mod_schechter}). The cyan line corresponds to the result of the HIPASS 21cm survey \citep{Zwaan2010}, while the red line represents the cumulative VF measured in the Local Volume by \citet{Klypin2014}. 
%% Conclusions: 
The two latter results are plotted to illustrate the magnitude of systematic uncertainties affecting the measurement of the VF.     
\textit{bottom panel:} The gray symbols represent our sample of galaxies with literature HI rotation curves (same as in Fig. \ref{fig:vrotvhalo+gals}). The colored lines represent the \vrotvhalo \ relation that corresponds to each of the cumulative VFs plotted in the upper panel, using the same color coding.
%% Conclusions: N/A yet
}
\label{fig:aa_vs_others}
\end{figure}

In Figure \ref{fig:aa_vs_others} we illustrate how uncertainties on the measurement of the VF can impact the analysis in this article: The blue shaded region in the top panel of the figure shows the statistical uncertainty in the measurement of the ALFALFA VF. More specifically, the plotted range in the cumulative VF is derived from the 1$\sigma$ parameter range of the modified Schechter fit to the differential ALFALFA VF (see Fig. \ref{fig:wf+vf}). 

We also plot in the same panel two recent literature determinations of the galactic VF, in order to illustrate the systematics affecting the measurement. First, we show the result obtained by the HI Parkes All Sky Survey (HIPASS), which is a blind, wide-area 21cm survey that predated ALFALFA \citep{Zwaan2010}. Second, we show the distribution measured within the Local Volume by \citet{Klypin2014}. The measurement is based on a nearly volume-complete catalog of galaxies within $D<10$ Mpc \citep{Karachentsev2013}, and so it also includes gas-poor systems that may be missing from an HI-selected sample.

The bottom panel of Figure \ref{fig:aa_vs_others} shows the resulting \vrotvhalo \ relations obtained from each of the cumulative VFs plotted in the upper panel. As we can clearly see by comparing the results of ALFALFA and HIPASS, a difference in the normalization of the VF leads to a horizontal shift in the inferred \vrotvhalo \ relation. Moreover, differences in the measured low-velocity slope of the VF affect the sharpness  of the relation's ``downturn''. This is evident by comparing the ALFALFA and Local Volume results. Overall, however, the bottom panel demonstrates  that random and systematic observational uncertainties on the inferred \vrotvhalo \ relation do not seem to strongly affect the conclusions reached in \S\ref{sec:results}.

\subsection{Low-velocity gas-poor galaxies?}
\label{sec:early_types}

The ALFALFA VF is not a complete census of galaxies in the universe, since objects with $M_{HI} < 10^7 \; M_\odot$ are not included in the measurement. As shown in \S\ref{sec:vel_am}, HI selection can complicate the measurement of the VF at the high-velocity end, where massive early-type galaxies dominate \citep[see also extensive discussion in][]{Obreschkow2013}. Similarly, we expect that HI selection will exclude a portion of the galactic population at low velocities, as well. Some of the excluded objects will be gas-poor early-type dwarfs, which are usually found in dense environments or as satellites of larger hosts \citep[e.g.,][]{Geha2012}. This population is not expected to be large, since the vast majority of dwarf galaxies correspond to star-forming systems with late-type morphologies (e.g., \citealp[Fig. 11]{Karachentsev2013}; \citealp[Fig. 15]{Baldry2012}). 
At the same time however, some fraction of late-type objects will also be excluded, simply because galaxies with low \vrot \ tend to have low gas masses.

\begin{figure}
\centering
\includegraphics[width=\linewidth]{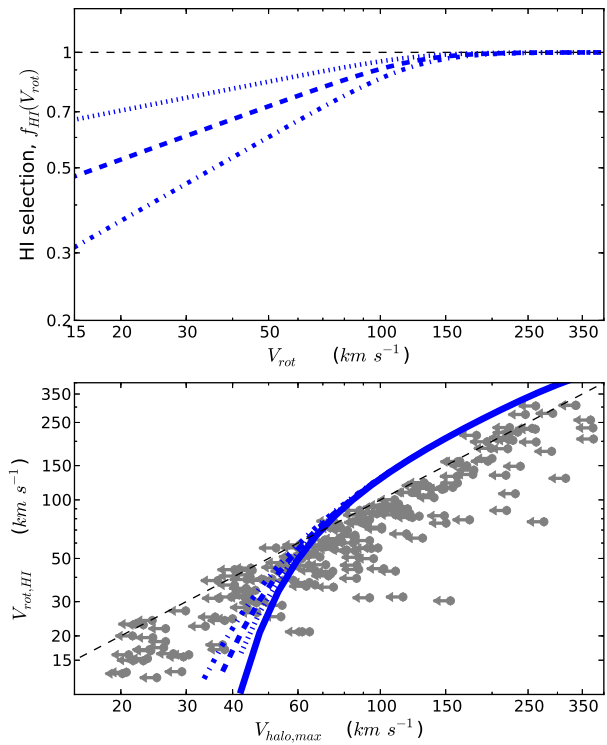}
\caption{  
%% title: 
Incompleteness of the ALFALFA survey at low velocities.
\textit{top panel:} The blue dotted, dashed and dash-dotted lines represent three relations for the ``HI-selection'' factor, $f_{HI}(V_{rot})$, each accounting for progressively more low-velocity systems that are undetectable by ALFALFA due to their low HI mass (Eqns. \ref{eqn:fhi1} - \ref{eqn:fhi3}). \textit{bottom panel:} The blue dotted, dashed and dash-dotted lines are the average \vrotvhalo \ relations for the corresponding \fhi$(V_{rot})$ relations shown in the top panel. For reference, we also plot our fiducial relation with a thick solid blue line (same as in Fig. \ref{fig:vrotvhalo}). The gray symbols represent our sample of galaxies with HI rotation curves, same as in Fig. \ref{fig:vrotvhalo+gals}. 
%%  Conclusions: The main conclusion of this work remains unchanged, even when a sizable population of low-velocity gas-poor galaxies is assumed.  
}
\label{fig:various_fhi}
\end{figure}

In order to explore this effect, we introduce here an ``HI-selection'' factor, \fhi$(V_{rot})$: it represents the fraction of galaxies at a given rotational velocity that are accounted for in the ALFALFA VF. We parametrize \fhi$(V_{rot})$ with the following analytical form:

\begin{equation}
f_{HI}(V_{rot}) = f_{HI,0} \times 2^{-\frac{\alpha}{\gamma}} \times \left( \frac{V_{rot}}{\tilde{V}} \right)^\alpha \times \left[ \frac{1}{2} + \frac{1}{2} \left(\frac{V_{rot}}{\tilde{V}}\right)^\gamma \right]^{-\frac{\alpha}{\gamma}} \;\; .
\end{equation}

\noindent
The equation above describes a power-law with exponent $\alpha$ at low velocities, that transitions to a constant value of $f_{HI,0}$ at high velocities. The parameter $\tilde{V}$ determines the location of the transition, while $\gamma$ controls its sharpness.

Since it is hard to observationally constrain \fhi$(V_{rot})$, we consider three cases that correspond to progressively larger populations of gas-poor galaxies at low velocities. They correspond to the following set of $[f_{HI,0}, \tilde{V}, \alpha, \gamma]$ parameters:

\begin{eqnarray}
\label{eqn:fhi1}
f^{(1)}_{\mathrm{HI}}(V_{rot})&:& \;\; [1.0,126 \, \mathrm{km}\mathrm{s}^{-1},0.19,5.] \\ 
\label{eqn:fhi2}
f^{(2)}_{\mathrm{HI}}(V_{rot})&:& \;\; [1.0,126 \, \mathrm{km}\mathrm{s}^{-1},0.35,5.]\\ 
f^{(3)}_{\mathrm{HI}}(V_{rot})&:& \;\; [1.0,126 \, \mathrm{km}\mathrm{s}^{-1},0.55,5.] \;\; ,  
\label{eqn:fhi3}
\end{eqnarray}  

\noindent
and they are shown in the top panel of Fig. \ref{fig:various_fhi}. The bottom panel of Fig. \ref{fig:various_fhi} shows the \vrotvhalo \ relations that result from each choice of \fhi \ (same linestyle coding). 
To derive each of the three AM relations, we have first boosted the differential ALFALFA VF by the corresponding HI-selection factor, $f_{HI}(V_{rot})^{-1} \times n(V_{rot})$. 
%For reference we also show with a thick solid blue line our fiducial \vrotvhalo \ relation. 

Figure \ref{fig:various_fhi} shows that in all cases the main conclusion drawn from Fig. \ref{fig:vrotvhalo+gals} does not change. This is true even for $f^{(3)}_{\mathrm{HI}}$, which predicts that 68\% of the galaxies at the lowest \vrot \ values should have too little HI to be included in the ALFALFA sample.

\subsection{Stochastic galaxy formation?}
\label{sec:stochasticity}

The derivation of the \vrotvhalo \ relation rests on two fundamental assumptions inherent in the AM process: (\textit{i}) that each (sub)halo hosts a detectable galaxy and (\textit{ii}) that the average relation between \vrot \ and \vhalo \ is monotonic. However, it is difficult to observationally assess whether these two assumptions are valid, especially at the mass scales of extreme dwarf galaxies.

\begin{figure}
\centering
\includegraphics[width=\linewidth]{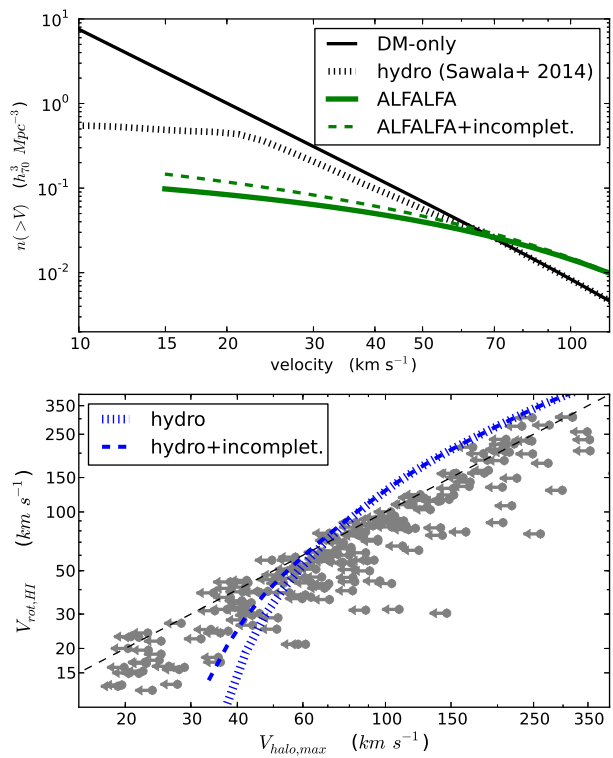}
\caption{
%% title: 
Baryonic effects on the number density of halos.
\textit{top panel:} The green line is the VF of all galaxies, as measured by ALFALFA (same as in Fig. \ref{fig:cum_vfs}). The green dashed line assumes an incompleteness of the ALFALFA VF at low velocities, parameterized as $f^{(2)}_{\mathrm{HI}}(V_{rot})$ (see \S\ref{sec:early_types}). The solid black line represents the VF of halos in the DM-only BolshoiP simulation (also same as in Fig. \ref{fig:cum_vfs}). The dotted black line corresponds to the HVF of halos hosting a stellar counterpart, according to the results of the hydrodynamical simulations of \citet{Sawala2014}. The flattening of the hydrodynamical HVF at velocities $\lesssim 25$ \kms \ is due to the suppression of galaxy formation caused by reionization feedback (see text for details). \textit{bottom panel:} The blue dotted line is the average \vrotvhalo \ relation according to the hydrodynamical result. The dashed blue line additionally takes into account a possible incompleteness of the ALFALFA VF. The gray symbols represent our sample of galaxies with resolved HI rotation curves. These datapoints are slightly shifted with respect to their positions in Fig. \ref{fig:vrotvhalo+gals}, because a correction for the cosmic baryon fraction is not necessary when comparing against a hydrodynamical simulation.%% Conclusions: Even though baryonic effects seem to somewhat alleviate the observed discrepancy at low velocities, the effect is too small to change the main conclusion of this article.
}
\label{fig:bent}
\end{figure}

In a recent article, \citet{Sawala2014} have explored this issue by means of hydrodynamical simulations. They find that DM-only halos with \vhalo $\lesssim 70$ \kms \ have larger total masses than their counterparts in more realistic simulations incorporating baryonic physics; they attribute this mass difference to the loss of baryonic material by moderately low-mass halos. Interestingly, they also find that among halos with \vhalo$ \lesssim 25$ \kms \ only a small fraction host a detectable stellar component in the hydrodynamic run. This steep decrease in galaxy formation efficiency for the lowest-mass halos is due to the effects of cosmic reionization. 
Based on these findings they argue that AM results based on DM-only simulations are not accurate at low velocities. 

In order to address these concerns, we re-derive here an average \vrotvhalo \ relation taking into account the baryonic effects described above. In particular, we use the cumulative HVF of halos that host a stellar counterpart in the hydrodynamic simulations of \citet{Sawala2014}. As shown by the top panel of Fig. \ref{fig:bent}, the hydrodynamic HVF deviates from the HVF of a DM-only simulation at \vhalo$\lesssim 70$ \kms, and then flattens out at \vhalo$\lesssim 25$ \kms. 

We then match the hydrodynamical HVF with the measured galactic VF measured by ALFALFA. The result is shown in the bottom panel of Fgi. \ref{fig:bent}: Even in the hydrodynamical case, the \vrotvhalo \ relation is not very different from the DM-only relation. This is because the number of halos exceeds the number of galaxies already at \vhalo $\approx 35$ \kms, in a regime where reionization is not yet effective (see top panel).  
As a result, the discrepancy between the AM relation and the internal kinematics of low-velocity dwarfs seems to persist, even when baryonic effects on the abundance of halos are considered.

The discrepancy is still present, even though somewhat alleviated, if one additionally assumes a substantial incompleteness of the ALFALFA VF (see bottom panel of Fig. \ref{fig:bent}). 
%This latter \vrotvhalo \ relation has been obtained by considering the ALFALFA VF corrected for incompleteness according to Eqn. \ref{eqn:fhi2} (green dashed line in top panel of Fig. \ref{fig:bent}). 
A similar result for the \vrotvhalo \ relation is also obtained when one considers the galactic VF measured by \citet{Klypin2014} in the Local Volume.

\subsection{Baryonic effects on dwarf galaxy rotation curves?}
\label{sec:baryon_rc}

Recent results from hydrodynamic simulations of galaxy formation have shown that repeated gas-blowout episodes, driven by bursty star-formation activity, can create a ``cored'' central DM profile in halos hosting dwarf galaxies (\citealp{Governato2012,PontzenGovernato2012}, but see also \citealp{delPopolo2014} for a different mechanism based on dynamical friction). As already discussed in Sec. \ref{sec:discussion}, \citet{Zolotov2012} argue that accounting for this effect is critical in the context of the satellite TBTF problem \citep[see also][]{BrooksZolotov2014}.

\begin{figure}
\centering
\includegraphics[width=\linewidth]{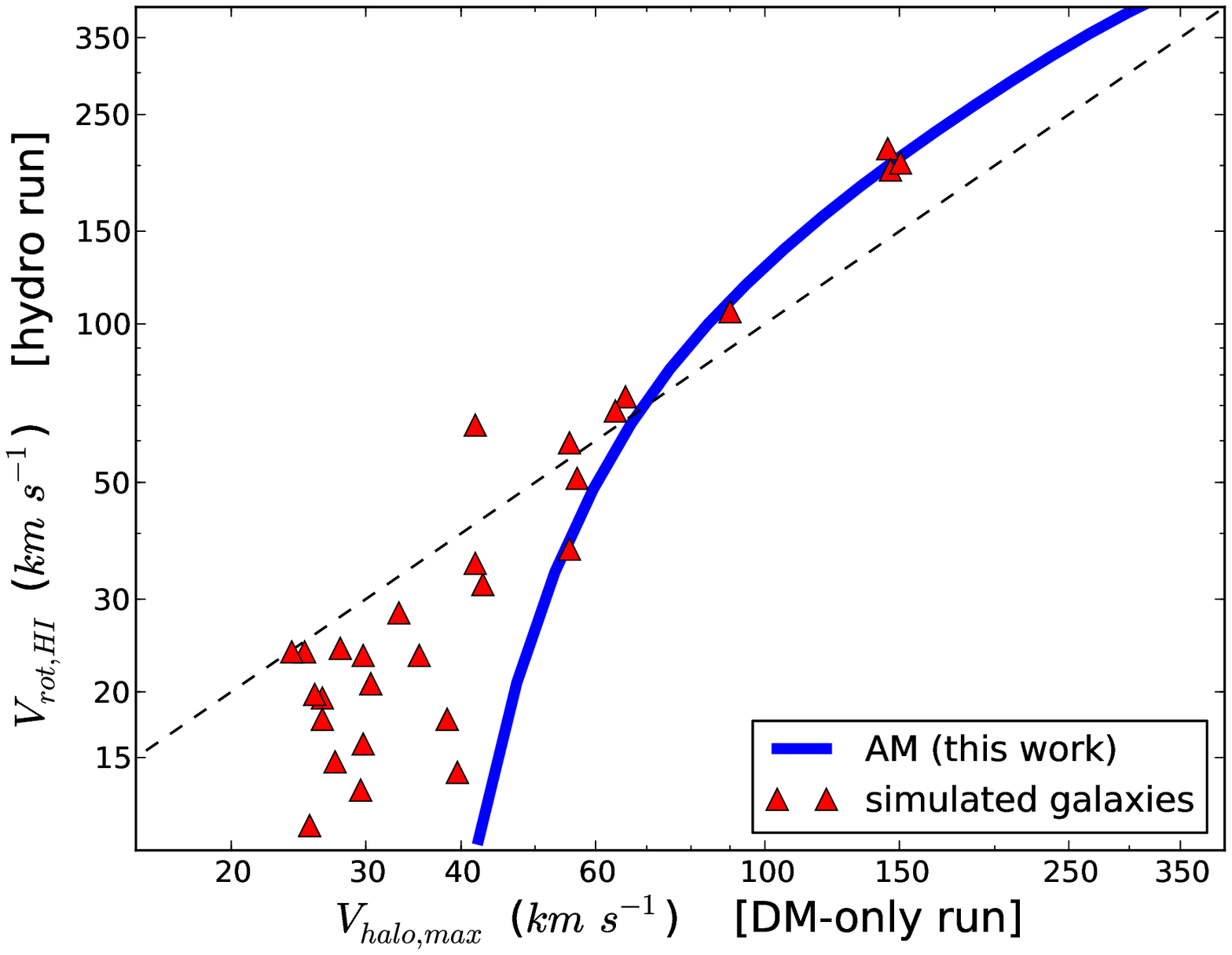}
\caption{
%% title: 
Baryonic effects on galactic rotation curves.
The blue solid line is the average \vrotvhalo \ relation in a \LCDM \ universe (same as in Fig. \ref{fig:vrotvhalo}). The red triangles represent galaxies from a set of hydrodynamic simulations which include efficient baryonic feedback \citep{Governato2012,BrooksZolotov2014,Christensen2014}. 
%% Conclusion:
Simulated low-velocity galaxies fall in the same region of the \vrotvhalo \ diagram as the actual dwarfs plotted in Fig. \ref{fig:vrotvhalo+gals}. This demonstrates that baryonic modifications of the galaxies' rotation curves do not significantly affect the simplified analysis performed in this work. }
\label{fig:brooks}
\end{figure}

Here we try to assess whether baryonic modifications to the RC of low-mass halos can affect the main result of this work. To this end, we place on the \vrotvhalo \ diagram 28 galaxies produced in an ensemble of hydrodynamical simulations \citep{Governato2012,BrooksZolotov2014,Christensen2014}; these are shown in Figure \ref{fig:brooks}. It is worth noting that the values of \vrot \ and \vhalo \ plotted here for the simulated objects have the same definition as for the AM analysis: \vrot \ is directly measured from simulated edge-on HI profiles (as per Fig. \ref{fig:profiles}), and \vhalo \ is the maximum circular velocity of the host halo in the \textit{DM-only} version of the simulation.

Low-velocity \textit{simulated} galaxies share the same locus on the \vrotvhalo \ diagram as the observed dwarf galaxies. This is a crucial point, because it shows that the simplified kinematic analysis performed in this article (based on NFW profiles motivated by DM-only simulations) agrees with the results of the hydrodynamic simulations (which include DM profile modification by baryonic feedback). In particular, simulated galaxies with \vrot$\lesssim 25$ \kms \ lie systematically to lower \vhalo \ values than predicted by the AM relation; this placement suggests that the number density of galaxies in the hydrodynamical simulation will be higher than what measured by ALFALFA. This statement is of course true only as long as the effects described in \S\ref{sec:uncertain_vf}-\ref{sec:stochasticity} are in the range explored in this work. 

Before we conclude this section we would like to issue a few cautionary notes. First, the comparison between actual and simulated galaxies is only valid provided that the latter have realistic HI properties. Most importantly, simulated galaxies should have HI disk sizes similar to the ones measured in actual dwarfs. However, due to the fact that the HI data for the Governato/Brooks/Christensen et al. simulation sets are not publicly available at present, this analysis is deferred for a future publication \citep{BrooksPapastergis2015}. Second, the main conclusions of this section are drawn based on the results of the specific simulation sets considered here. We cannot therefore exclude the possibility that different simulations --or even the same simulations carried out with different feedback prescriptions-- may give substantially different results \citep[for example, cf.][]{BrookdiCintio2014}.

\subsection{Alternative dark matter models?}
\label{sec:wdm}

Besides baryonic effects, several alternative DM models have been considered to provide solutions to the small-scale challenges faced by \LCDM. Perhaps the most well studied among them is the warm dark matter (WDM) model, characterized by a DM particle with mass in the $\sim $keV range. Such ``light'' particles (compared to $m_{CDM} \sim 10 \; \mathrm{GeV} - 1 \; \mathrm{TeV}$) result in a suppression of structure on spatial scales that are relevant for galaxy formation \citep[e.g.,][]{Zavala2009,Menci2012}. This property of WDM has been regarded as a natural way to resolve CDM overabundance issues. For example, several authors have argued that a WDM model with $m_{WDM} \approx 1$ keV could plausibly reproduce the flatness of the velocity function  \citep[e.g.,][]{Zavala2009,Zwaan2010,Papastergis2011}. In addition, a WDM model with $m_{WDM} = 1.5-2$ keV could potentially provide a solution to the satellite TBTF problem \citep[e.g.,][]{Lovell2012,Schneider2014}; this is because MW-sized halos in WDM have fewer --and less dense-- massive subhalos compared to CDM.

\begin{figure*}
\centering
\includegraphics[width=\linewidth]{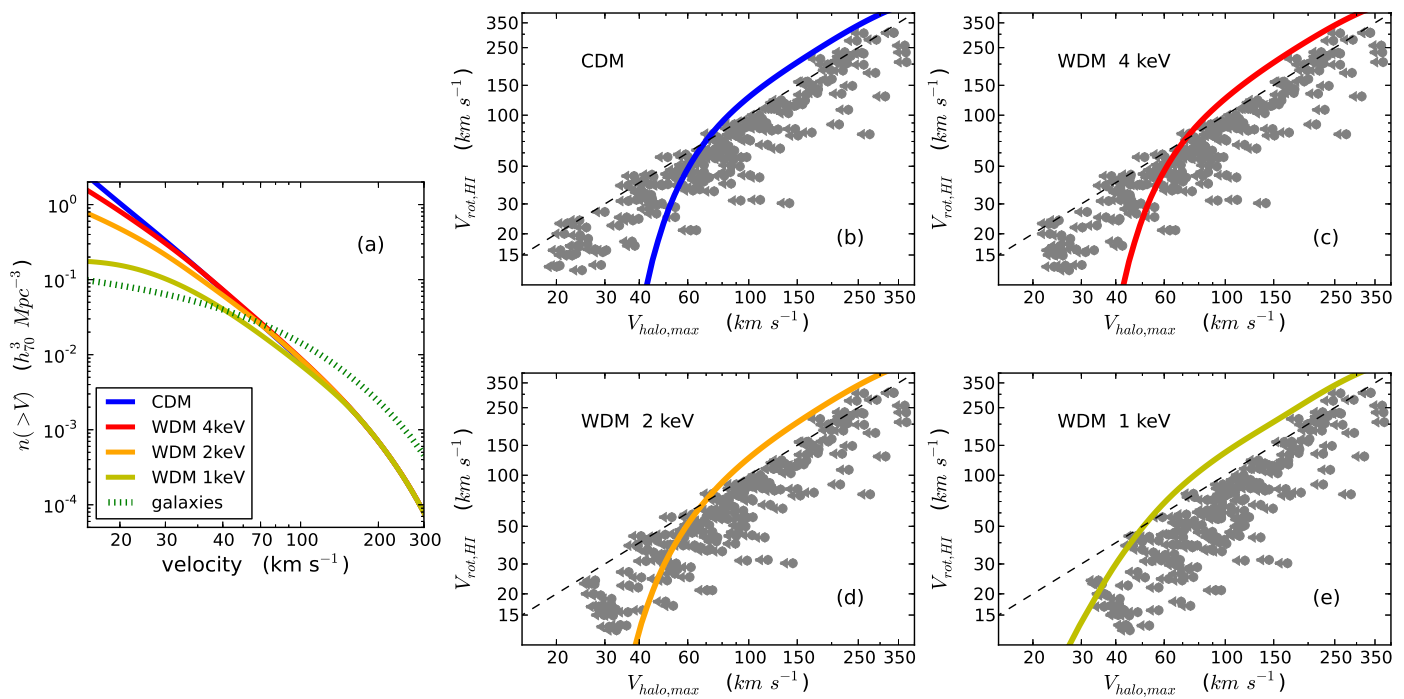}
\caption{
%% title: 
\vrotvhalo \ diagrams in WDM cosmologies.
\textit{panel a:} The blue solid line corresponds to the cumulative halo velocity function in a CDM model (same as black solid line in Fig. \ref{fig:cum_vfs}). The red, orange and yellow solid lines correspond to WDM models with particle masses of 4keV, 2keV and 1keV, respectively \citep{Schneider2014}. The dotted green line is the cumulative VF for galaxies of all morphological types (same as solid green line in Fig. \ref{fig:cum_vfs}). \textit{panels b-e:} The colored solid lines correspond to the \vrotvhalo \ relations in each dark matter model (same color coding as in panel \textit{a}), inferred by AM. The gray datapoints represent our sample of literature galaxies with HI rotation curves. The same galaxy can be assigned a different value of \vhalo \ in each of the four panels, because each model has a different median mass-concentration relation \citep{Schneider2012}. 
%% Conclusions: In text.
}
\label{fig:wdm}
\end{figure*}

Here we assess whether WDM can also provide a solution to the field TBTF problem. Panel \textit{a} of Figure \ref{fig:wdm} compares the cumulative HVF of CDM with the corresponding distributions for three WDM models, with $m_{WDM} = 1,2 \: \& \: 4$ keV. The halo counts for the WDM models have been obtained from the simulations of \citet{Schneider2014}. The figure clearly shows that, as $m_{WDM}$ decreases, the number density of low-velocity halos becomes increasingly more suppressed with respect to CDM. The difference in the cumulative HVFs translates then into different inferred \vrotvhalo \ relations for each model; the latter are shown in panels \textit{b}-\textit{e} of Fig. \ref{fig:wdm}. At the same time, the typical concentration of halos in WDM models is lower than in CDM. In order to place galaxies on the \vrotvhalo \ diagram for each WDM model, we repeat the process described in \S\ref{sec:kinematic_analysis} using each time the appropriate median $c$-$M_{vir}$ relation \citep[][Eqn. 39]{Schneider2012}. Generally, a given galactic RC can accommodate a progressively more massive host halo, as $m_{WDM}$ decreases.

Panel \textit{e} shows that the AM relation predicted by the 1keV model seems to be fully consistent\footnotemark{} with the upper limits derived from individual galaxies. The reason is that both WDM effects discussed above (suppression of the number of low-mass halos \& lower halo concentrations) are very pronounced in this model. On the other hand, these same effects are much less prominent in the 2keV case, even though they are still clearly discernible. Panel \textit{d} shows that low-velocity galaxies are not fully consistent with the predicted \vrotvhalo \ relation, even for a model with a particle mass as low as $m_{WDM} = 2$keV. Nonetheless, the inconsistency is significantly alleviated in this case, and perhaps accounting for any of the uncertainties in \S \ref{sec:uncertain_vf}-\ref{sec:baryon_rc} could be sufficient to fully resolve the tension.

\footnotetext{Since the kinematics of galaxies are only used to set conservative upper limits on \Vhalo, consistency between the AM relation and the individual datapoints does not guarantee that a model can reproduce the observed VF. For example, \citet{Schneider2014} \& \citet{Klypin2014} have argued that WDM cannot reproduce in detail the measured galactic VF, regardless of $m_{WDM}$.}

Despite their promise, WDM models with particle masses in the range 1-2 keV are at odds with a number of independent constraints (at least when interpreted as thermal relics). For example, measurements of the small-scale power of the Ly-$\alpha$ forest at high-$z$ place a 2$\sigma$ lower limit of $m_{WDM} > 3.3$ keV \citep{Viel2013}. In addition, WDM models with $m_{WDM} < 2.3$ keV are ruled out by the number of ultra-faint satellites observed around the MW \citep[e.g.,][]{PolisenskiRicotti2011}. As panel \textit{c} shows, WDM models that do not violate the constraints mentioned above are practically indistinguishable from CDM. For example, in the 4keV case galaxies with \vrot$\lesssim 25$ \kms \ are clearly incompatible with the predicted AM relation. We therefore conclude that it is unlikely that (thermal relic) WDM can provide a solution to the field TBTF problem, without violating other independent astrophysical constraints.

\section{Summary}
\label{sec:summary}

We measure the distribution of galactic rotational velocities in the nearby universe, using the dataset of the ALFALFA 21cm survey. Based on the measured velocity function (VF), we statistically connect galaxies with their host halos via the technique of abundance matching (AM). In a \LCDM \ universe, linewidth-derived rotational velocities (\Vrot) of dwarf galaxies are expected to underestimate the maximum rotational velocity of their host halos (\Vhalo). For example, galaxies with \Vrot$\approx 15$ \kms \ are expected to be hosted by halos with \Vhalo$\gtrsim 40$ \kms \ (see Fig. \ref{fig:vrotvhalo}). This trend reflects the fact that, at the low end, the halo velocity function (HVF) in \LCDM \ rises much faster than the observed galactic VF (see Fig. \ref{fig:cum_vfs}).

We then compile an up-to-date literature sample of galaxies with HI rotation curves, to observationally test the predicted \vrotvhalo \ relation. Our sample contains a large number of extremely low-velocity dwarfs, mainly drawn from the FIGGS, SHIELD and LITTLE THINGS samples. For each galaxy we find the most massive NFW halo that is compatible with the last measured point (LMP) of the galactic rotation curve. Galaxies can then be placed on the \vrotvhalo \ diagram as upper limits in \vhalo \ (see Fig. \ref{fig:find_vhalo}). As Figure \ref{fig:vrotvhalo+gals} shows, the upper limits derived from individual galaxies are consistent with the average AM relation, for most of the range in galactic rotational velocities ($30 \; \mathrm{km}\,\mathrm{s}^{-1} \lesssim V_{rot} \lesssim 300 \; \mathrm{km}\,\mathrm{s}^{-1}$). Most importantly however, this is not the case for the lowest velocities probed: for example, at \Vrot$\lesssim 25$ \kms \ the HI rotation curves of galaxies cannot accommodate host halos as massive as predicted by \LCDM. This work therefore confirms the similar results previously found for field galaxies by \citet{Ferrero2012} and for non-satellite galaxies in the Local Group by \citet{Kirby2014} \& \citet{Garrison2014}. At the same time, the analysis performed in this work addresses several caveats present in these previous studies (see Sec. \ref{sec:discussion}).

The discrepancy described above is directly analogous to the too big to fail problem (TBTF) faced by the bright satellites of the Milky Way, but here is observed for galaxies in the field. This finding has therefore important implications, because several of the proposed solutions to the satellite TBTF problem are not applicable in this case. For example, the galaxy population studied in this work mainly consists of gas-rich, fairly isolated objects, that have not been heavily affected by processes such as tidal stripping. As a result, potential solutions to the TBTF problem that rely on strong environmental effects \citep[e.g.,][]{Zolotov2012} cannot explain the presence of a similar discrepancy in the field.

We furthermore consider a number of assumptions and sources of uncertainty that may impact the main conclusions of this work. These include, for example, observational uncertainties on the measurement of the VF and effects related to the bias of HI surveys against gas-poor galaxies. Perhaps the two most important among them, however, are baryonic modifications to the abundance of galaxy-hosting halos and to the velocity profiles of dwarf galaxies. We show that the former baryonic effect does not seem to be able to resolve the reported discrepancy, at least when considered on its own (\S\ref{sec:stochasticity}). Furthermore, a preliminary analysis of a set of hydrodynamical simulations \citep{Governato2012,BrooksZolotov2014,Christensen2014} suggests that the latter effect is not able to resolve the discrepancy, as well (\S\ref{sec:baryon_rc}). At the same time, it is still possible that combinations of several of the effects mentioned above may prove sufficient to provide a solution. It is also possible that the baryonic effects considered here may be larger than indicated by the theoretical models used in this work. 

Lastly, we check whether an alternative warm dark matter (WDM) model can provide a solution to the field TBTF problem. We find that the inconsistency is lifted in WDM models with particle masses of $\approx$1 keV. However, (thermal relic) particles with such light masses are in conflict with a number of independent observational constraints (see \S\ref{sec:wdm} for details).

\begin{acknowledgements}
The authors would like to thank an anonymous referee for their constructive comments and the language editor for their thorough work. The authors would particularly like to thank John Cannon, SeHeon Oh and Alyson Brooks for sharing their yet-to-be-published data. Erwin de Blok, John Cannon, Alyson Brooks, Thijs van der Hulst, Mike Boylan-Kolchin, Marla Geha, Tjitske Starkenburg and Mike Jones provided insightful feedback and comments. EP would also like to thank the Aspen Center for Physics for its hospitality (ACP is supported by NSF grant 1066293), and Mary Gooding for her valuable advice. \\ 
EP is supported by a NOVA postdoctoral fellowship at the Kapteyn Institute. The ALFALFA team at Cornell is supported by U.S. NSF grant AST-1107390 to MPH and RG and by grants from the Brinson Foundation.
\end{acknowledgements}

\appendix

\section{Details on the measurement of the velocity function}
\label{sec:appendix_a}

In order to measure galactic number densities as accurately as possible, we apply well-defined selection rules and quality cuts to the \aforty \ sample. In particular:

\begin{itemize}

\item Sources are selected from two rectangular sky regions of the \aforty \ footprint, which span the same RA range of $08^h00^m < \alpha < 16^h20^m$ and the Dec ranges $4^\circ < \delta < 16^\circ$ \& $24^\circ < \delta < 28^\circ$, respectively. Both regions are located in the northern Galactic cap, and have almost complete overlap with the SDSS DR7 footprint.

\item Only confidently detected ALFALFA sources that are classified as extragalactic objects are considered. In particular, only sources designated as ``Code 1'' in \aforty \ and detected at $(S/N)_{HI} > 6.5$ are included in the sample. 

\item Only ALFALFA sources that are crossmatched with a photometric object in the SDSS DR7 are selected. This criterion aims at removing HI sources that are unlikely to be hosted by their own individual halo, such as tidal HI clouds or diffuse gas found in the vicinity of interacting systems. Bona fide galaxies detected by ALFALFA that lack a photometric counterpart in SDSS images are instead extremely rare\footnotemark{} (see \S4.3 in \citealp{Haynes2011}).    

\footnotetext{This statement does not apply to the Local Group and its immediate surroundings. In particular, the dwarf galaxy LeoP has been recently discovered by its 21cm emission \citep{Giovanelli2013}, and several more candidate low-mass galaxies have been identified in the same way \citep{Adams2013}.}

\item A minimum distance cutoff is imposed at $D_{min} = 7$ Mpc. This minimum distance limit is imposed in order to exclude from the sample objects that have a very large uncertainty on their distance, and can therefore introduce large errors in the measurement. At the same time however, this cutoff will exclude some genuinely very low-mass galaxies that can only be detected locally. The distances adopted here are the ones listed in the \aforty \ catalog, which are primarily based on the flow model of \citet{Masters2005} (for details refer to \S3.2 in \citealp{Martin2010}). Keep in mind that flow model distances for very nearby objects can have large fractional errors \citep[see e.g.,][]{McQuinn2014}. In general, distance uncertainties result in an overestimate of the number density of low-velocity galaxies, even though the effect is not large \citep[\S4.2]{Papastergis2011}.    

A maximum distance cutoff is defined in terms of recessional velocity in the CMB frame of reference, as $cz_{max} = 15\,000$ \kms \ ($D_{max} \approx 214$ Mpc). This maximum recessional velocity limit is imposed in order to avoid a spectral region that is heavily affected by radio frequency interference. 

\item HI sources that lie below the 50\% completeness limit of the ALFALFA survey are excluded from the sample. The completeness limit of the ALFALFA survey has been measured in Section 6 of \citet{Haynes2011}. In this work we use Eqns. 4 \&5 in \citet{Haynes2011}, to define the flux values corresponding to 50\% completeness as a function of velocity width\footnotemark{}.

\footnotetext{In principle, the full two-dimensional surface of completeness in the flux-width plane is available in Section 6 of \citet{Haynes2011}. In this work, we use a step function approximation to this surface, whereby the completeness is assumed to be 1 above the 50\% completeness flux and 0 below. \citet{RosenbergSchneider2002} have argued that such an approximation produces fairly accurate results, but the more detailed analysis of \citet{Obreschkow2013} warns about possible biases.}  

\item We restrict our sample to the HI mass range of $M_{HI} = 10^7 - 10^{11} M_\odot$ and the velocity width range of $ W_{50} = 28 - 900$ \kms. These lower bounds on HI mass and velocity width are conservative, and do not reflect the performance limitations of the ALFALFA survey. 

\end{itemize}

\begin{figure*}
\centering
\includegraphics[scale=0.46]{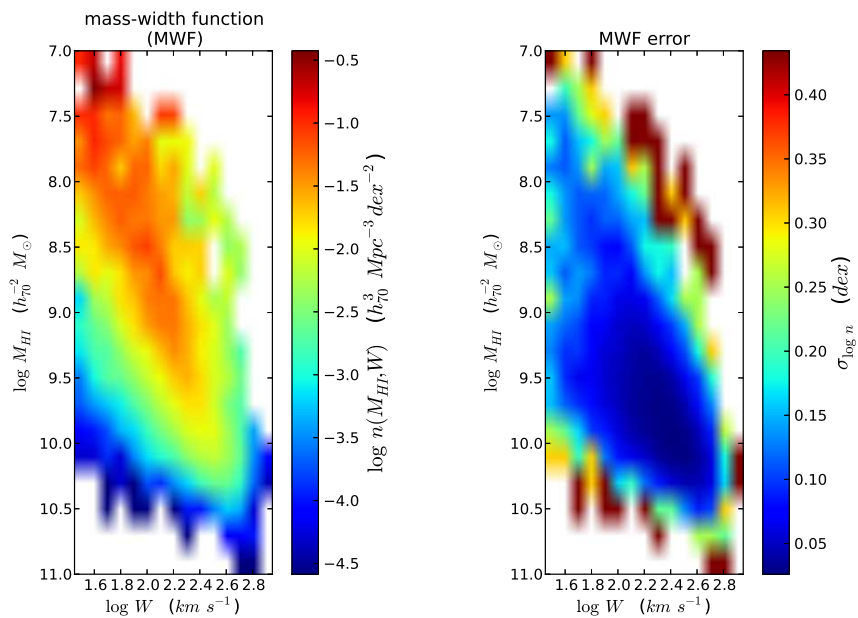}
\caption{ 
%% title: 
ALFALFA mass-width function.
\textit{left panel:} The colormap represents the mass-width function (MWF) measured by ALFALFA, i.e., the number density of galaxies within logarithmic bins of HI mass (\mhi) and velocity width ($W$). The measurement spans the range $M_{HI} = 10^7 - 10^{11} \; M_\odot$ and $W = 28 - 900$ \kms. Note that the stretch of the colormap is logarithmic. \textit{right panel:} Colormap of the 1$\sigma$ statistical error on the number density, expressed in dex (i.e., $\sigma_{\log\,n} = \frac{\sigma_n / n}{\ln(10)}$). The plotted errors are just due to counting statistics, and  systematic uncertainties are not included. 
%% conclusions: N/A
}
\label{fig:mwf}
\end{figure*}

The above set of selection rules results in a final sample of $6\,770$ sources, located over $\approx 2\,900$ deg$^2$ of sky and within a volume of approximately $2\times10^6$ Mpc$^3$. This is the sample used to calculate the mass-width function (MWF) of galaxies, shown in the left panel of Figure \ref{fig:mwf}. The horizontal axis of the panel is the rest frame velocity width, $W$, which is derived from the observed velocity width, $W_{50}$, after a Doppler broadening correction is applied: $W = W_{50}/(1+z_\odot)$. The vertical axis is HI mass, \mhi, calculated from a source's HI flux, $S_{HI}$, and distance, $D$, through the relation $M_{HI}(M_\odot) = 235.6 \cdot S_{HI}(\mathrm{mJy}\:\mathrm{km}\,\mathrm{s}^{-1}) \cdot D^2(\mathrm{Mpc})$. The color scale shows the number density of galaxies within logarithmic bins in $M_{HI}$ and $W$, i.e., $n(M_{HI},W) = \frac{dn_{gal}}{d\log(M_{HI}) \, d\log(W)}$. 
%Note that the colorbar stretch is logarithmic (i.e., $\log n$ is plotted). 
The right panel of Fig. \ref{fig:mwf} shows the error on the number density, expressed in dex (i.e., $\sigma_{\log\,n} = \frac{\sigma_n / n}{\ln(10)}$). The error plotted here only includes the uncertainty from counting statistics, and is therefore lowest in the mass-width bins with the highest number of detections (and vice-versa). Several other sources of  uncertainty, most of them systematic, are not included in the errors above (see however discussion in Sec. \ref{sec:uncertain_vf}).

Since the ALFALFA sample is not volume-limited, the MWF shown in Fig. \ref{fig:mwf} is not just a simple number count of detections normalized by the survey volume. In particular, galaxies with different masses and widths can be detected over different volumes, so each source has to be weighted by an appropriate volume-correction factor. In this work, we calculate weighting factors via the ``$1/V_{eff}$'' maximum-likelihood method. A thorough description of the method and full details of its implementation can be found in \citet{Papastergis2011} and \citet{Zwaan2010} and references therein, but here we very briefly summarize its most important characteristics: For each galaxy $i$ in the final sample we calculate an ``effective'' volume, $V_{eff,i}$, which depends on the object's HI mass, velocity width and distance ($M_{HI,i}$, $W_{50,i}$ \& $D_i$, respectively). The value of the MWF in mass bin $j$ and width bin $k$ and its corresponding ``counting'' error can then be calculated as 

\begin{eqnarray}
n_{jk} & = & \frac{1}{\Delta m_{HI}\, \Delta w} \; \sum_i \frac{1}{V_{eff,i}} \;\;\;\; \mathrm{and}  \\
\sigma_{n_{jk}}^2 & = & \frac{1}{(\Delta m_{HI}\, \Delta w)^2} \; \sum_i \frac{1}{V_{eff,i}^2} \;\;,
\label{eqn:Veff}
\end{eqnarray}

\noindent
where the summation runs over all galaxies $i$ which belong to the specific bin. In the equations above, $\Delta m_{HI}$ and $\Delta w$ are the sizes of the logarithmic bins in mass and width (i.e., $m_{HI} = \log(M_{HI}/M_\odot)$ and $w = \log(W/\mathrm{km}\,\mathrm{s}^{-1})$). The equations above have the same form as the equations used in the standard ``$1/V_{max}$'' method. In the $1/V_{max}$ case, the quantity entering Eqn. \ref{eqn:Veff} would be $V_{max,i}$, which is defined as the maximum volume over which galaxy $i$ can be detected according to the survey sensitivity. In fact, $V_{eff,i}$ exactly coincides with $V_{max,i}$ for a spatially homogeneous sample. The advantage of the $1/V_{eff}$ method is that it takes into account density fluctuations in the survey volume, and therefore the resulting estimates are less susceptible to biases introduced by the presence of large-scale structure.

If we marginalize the two-dimensional MWF along each axis in turn, we obtain the HI mass function (HIMF) and the velocity width function (WF) of galaxies. These distributions are of central importance in extragalactic astronomy, and their measurement and scientific interpretation has been the subject of a large body of literature (see e.g., \citealp{Martin2010, Zwaan2005, Papastergis2011, Zwaan2010} for some of the most recent observational results).

\section{Last measured point (LMP) radii and velocities for dwarf galaxy sample}
\label{sec:appendix_b}

\begin{figure*}
\centering
\includegraphics[scale=0.53]{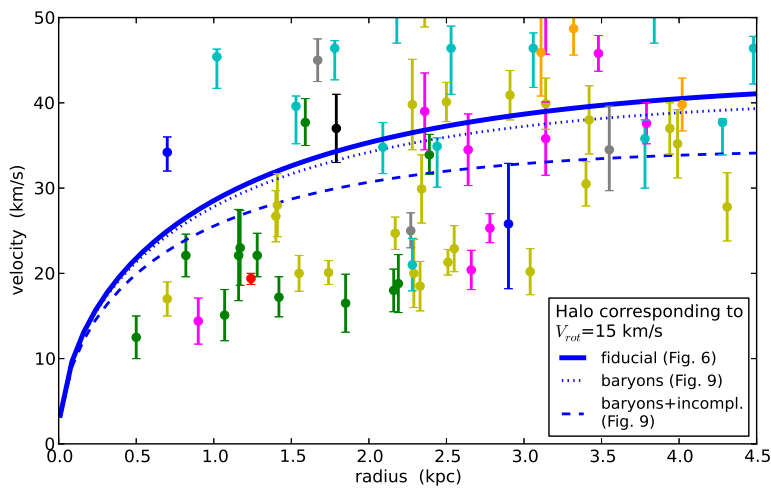}
\caption{ 
%% title: 
Last measured point radii and velocities for dwarf galaxies.
The datapoints represent the radii and velocities at the LMP for some of the galaxies with resolved HI observations (\S\ref{sec:data_sample}). This plot is restricted to $V_\mathrm{LMP} < 50$ \kms, where most of the ``failing'' extreme dwarf galaxies in Fig. \ref{fig:vrotvhalo+gals} are found. The datapoints are color coded according to their original data reference, same as in Fig. \ref{fig:vrotvhalo+gals}. The three blue lines represent the RCs of NFW halos that are assigned through AM to galaxies with \vrot$= 15$ \kms. In particular, the solid line represents the fiducial AM result (Fig. \ref{fig:vrotvhalo+gals}), the dotted line represents the AM result including baryonic effects on the abundance of halos, and the dashed line represents the AM result including baryonic effects on the abundance of halos and observational incompleteness of the ALFALFA VF (see Fig. \ref{fig:bent}).
%% conclusions: N/A
}
\label{fig:vout_radii}
\end{figure*}

Figure \ref{fig:vout_radii} shows the radii and velocities at the LMP for a subset of the interferometric galaxy sample described in \S\ref{sec:data_sample}. In particular, the figure zooms on the range $R_\mathrm{LMP} < 4.5$ kpc and $V_\mathrm{LMP} < 50$ \kms, where most of the ``failing'' dwarfs of Fig. \ref{fig:vrotvhalo+gals} reside. 
%The colors of the datapoints denote their original reference in the literature, following the same color scheme as in Fig. \ref{fig:vrotvhalo+gals}. 
Where appropriate, the plotted $V_\mathrm{LMP}$ values have been corrected for the effects of turbulent motions (see \S\ref{sec:results}).

For reference, we also plot the RC of halos that are assigned through AM to a galaxy with \vrot$= 15$ \kms. In particular, the solid RC represents our fiducial AM analysis (\S\ref{sec:vel_am} and Fig. \ref{fig:vrotvhalo+gals}), the dotted RC represents the AM analysis including baryonic effects on the abundance of halos (\S\ref{sec:stochasticity} and Fig. \ref{fig:bent}), and lastly the dashed RC refers to the AM analysis including baryonic effects and observational incompleteness of the ALFALFA VF (also \S\ref{sec:stochasticity} and Fig. \ref{fig:bent}). We note that the solid RC has been rescaled down according to the cosmic baryon fraction (see discussion in \S\ref{sec:kinematic_analysis}); as a result, the difference between the solid and dotted RCs reflects baryonic effects on the abundances of halos that are in addition to a simple rescaling.

Since 15 \kms \ is approximately the lowest value of \vrot \ encountered in our sample, these halos are effectively the smallest halos that are expected to host galaxies according to each AM result. Consequently, galaxies whose LMP measurement lies below each of the plotted RCs, are (almost certainly) a ``failure'' in Fig. \ref{fig:vrotvhalo+gals}. At the same time, Fig. \ref{fig:vout_radii} shows that most failing dwarfs have $R_\mathrm{LMP} \lesssim 3$ kpc. As a result, baryonic modifications to the inner velocity profiles of halos should be taken into account when interpreting the results of this article. Fig. \ref{fig:vout_radii} also illustrates the point that baryonic modifications to the RCs of halos can only resolve the issue if baryonic feedback can have a sizable impact at radii as large as 2-3 kpc. However, the preliminary analysis of hydrodynamic simulations described in \S\ref{sec:baryon_rc} suggests that this is not the case (but keep also in mind cautionary notes in \S\ref{sec:baryon_rc}).

\end{document}